\documentclass[aps,pre,twocolumn,groupedaddress,showpacs]{revtex4-1}
\usepackage[final]{graphicx}
\usepackage{amsmath}
\usepackage{amssymb}
\usepackage{amsthm}
\usepackage{mathtools}
\usepackage{graphicx} 
\usepackage{epsfig}
\usepackage{ifthen}
\usepackage[utf8]{inputenc}
\usepackage[english]{babel}
\usepackage[T2A]{fontenc}
\usepackage{setspace}
\usepackage{url}
\usepackage[svgnames]{xcolor}
\usepackage{hyperref}
\usepackage{stackrel}
\usepackage{afterpage}
\usepackage[misc]{ifsym}

\def\d{d}

\def\const{\operatorname {const}}
\def\sign{\operatorname {sign}}

\def\sign{\operatorname{sign}}
\def\const{\operatorname{const}}

\newtheorem{remark}{\it Remark}

\def\I{\mathrm i}

\def\cc{\mathrm{c.c.}}

\renewcommand{\=}{\stackrel{\mbox{\scriptsize def}}{=}}

\def\pFO{}
\def\tphi{\psi}

\def\EXP#1{\mathrm e^{#1}}

\def\VA{{v}}
\let\mathscr=\mathcal

\def\de{\delta}

\def\DM{m-1}

\newcommand{\minfty}{m_\infty}
\newcommand{\Kinfty}{K_\infty}
\newcommand{\Binfty}{B_\infty}

\begin{document}
\selectlanguage{english}

\title{Anti-localization of non-stationary quasi-waves in a strongly nonlinear $\beta$-FPUT chain%
}

\author{Serge N. Gavrilov}
\affiliation{Institute for Problems in Mechanical Engineering RAS, St.~Petersburg, Russia}
\email{serge@pdmi.ras.ru}
\author{Ekaterina V. Shishkina}
\email{shishkina\_k@mail.ru}
\affiliation{Institute for Problems in Mechanical Engineering RAS, St.~Petersburg, Russia}
\author{Bogdan S. Borisenkov}
\email{nickendsm@gmail.com}
\affiliation{Institute for Problems in Mechanical Engineering RAS, St.~Petersburg, Russia}
\affiliation{Peter the Great St.~Petersburg Polytechnic University (SPBPU), St.~Petersburg, Russia}


\begin{abstract}
  Recently,
  a new general wave phenomenon, namely
  ``the anti-localization of non-stationary linear waves'', 
  has been introduced and discussed 
  (Shishkina et al., J. Sound. Vib. 553, 2023, 117673). This is zeroing 
  of the propagating component for a non-stationary wave-field near
  a defect in infinitely long wave-guides. The phenomenon is known to be observed in both
  continuum and discrete mechanical systems with a defect, provided that the frequency spectrum 
  for the corresponding homogeneous system possesses a stop-band. 
  In this paper, we show that the anti-localization is also quite common for
  nonlinear systems. To demonstrate this, we numerically solve several
  non-stationary problems for an infinite strongly nonlinear $\beta$-FPUT chain with a defect.  
  In our opinion, the anti-localization essentially influences the processes 
  of heat transfer in linear and nonlinear lattices.
\end{abstract}

\maketitle


\section{Introduction}

In infinitely long uniform linear continuum and discrete systems, the perturbations caused by a pulse point load
travel as waves \footnote{For a discrete system, it is more correct to speak about quasi-waves.}
along a system from the load application point to infinity. The
presence of an inhomogeneity (or a defect) at the load application point can
lead to accumulation or decumulation of the wave energy near the defect compared to the uniform case.
Apparently, the first study where this fact was demonstrated is the classical paper by Lamb 
\cite{Lamb1900}, where it was shown that attaching a discrete mass-spring system into a
continuum medium described by the 1D wave equation leads to the leakage of energy
of a mass-spring system. On the other hand, such an attaching
leads to the accumulation of energy in the wave-guide near the inhomogeneity,
since in
a uniform medium, described by the 1D wave equation, the perturbations caused by a point pulse load 
immediately leave the neighborhood of the application point, leaving it in
the rest \footnote{This is due to the structure of the fundamental solution for the 1D wave equation \cite{Vladimirov1971,Watanabe2015}.}.

In uniform linear dispersive systems, the amplitude of perturbations caused by a point
pulse load vanishes with time not immediately but in a continuous manner. Generally, in 1D systems the order of decay
for perturbations (e.g., for the particle velocities, the strains, or the displacements if the
system is fixed somewhere and does not have rigid modes of motion) is $1/\sqrt t$,
where $t\to\infty$ is time. This
result is well-known; see, e.g., \cite{Whitham1999,Slepyan1972}, and can be obtained by the asymptotic 
method of stationary phase
applied to a representation of the exact non-stationary solution in the
form of a Fourier integral. Again, introducing 
the inhomogeneity can
lead to the accumulation or decumulation of the wave energy nearby. 

The effect of
the accumulation can be related to the so-called trapped modes \cite{ursell1951trapping,Ind-book-R2E,Kuznetsov2002,Andrianov2013,Gavrilov2024nody}, which can
exist in dispersive systems. In infinite discrete systems (lattices or chains), the analogous
phenomenon is known as the impurity-induced localized modes
\cite{Montroll1955,Kivshar1991,Kivshar1997,Mamalui1999}. 
The existence of such modes leads to trapping of some portions of the wave
energy of a non-stationary process forever near a load application point 
at an inhomogeneity \cite{kaplunov1986torsional,Teramoto1960} (we call this the localization of non-stationary waves). The
whole frequency spectrum of our system can have a stop-band, where complex
wave-numbers correspond to real frequencies. In such systems, so-called strong
type of localization \cite{luongo2001mode,Luongo1992} is quite common, which is related to a pole in the
denominator of the system Green function in the frequency domain lying inside the stop-band.
The rest of the energy (or all the wave energy if there is no localization) still travels
to infinity as a vanishing propagating component of the wave-field,
where perturbations again have the order of decay $1/\sqrt t$ (in a 1D system).

Recently \cite{Shishkina2023cmat,Shishkina2023jsv,Gavrilov2024nody}, an opposite general wave phenomenon, 
which has been called {\it the anti-localization of non-stationary waves}, 
related to the decumulation of the energy near an 
inhomogeneity, has been introduced.
It is observed in both discrete and continuum infinite linear systems, having
the dispersion graph with a stop-band \footnote{The frequency spectrum of a discrete linear infinite system always has a stop-band
\cite{Kossevich1999}
for high values of the frequency.}.
The presence of a stop-band leads to the existing of the cut-off frequency to which
zero group velocity corresponds. The oscillation corresponding to the cut-off frequency
accumulates near a point pulse source \cite{hemmer1959dynamic,slepyan1987energy}; this can cause the phenomenon of the
anti-localization. 
The existence of such a phenomenon is in agreement with some particular 
results previously obtained 
\cite{kaplunov1986torsional,hemmer1959dynamic,Mueller1962,Mueller2012,Kashiwamura1962,Rubin1963,Jackson1978}
for some systems, where the strong localization can take place.
The anti-localization is zeroing of the leading-order term of order 
$1/\sqrt t$ in asymptotic expansions for perturbations at an inhomogeneity. Consequently, 
the next term of order $1/\sqrt {t^3}$
becomes a leading-order term there.
Accordingly, the propagating perturbations,
in an expanding with time neighborhood of the inhomogeneity, become asymptotically small 
compared to ones in zones closer to the leading wave-fronts. The amplitude of
oscillation in the propagating component has a clear almost zero minimum at the
inhomogeneity. The anti-localization can be observed in the
absence of the localization as well as co-exist with the localization.  In
the latter case, after some time, a peak of the amplitude for oscillation also
becomes clearly recognizable, being observed in the center of a zone where
the perturbation amplitude is almost zero.
Note that the term ``weak anti-localization'' is commonly known in modern physics \cite{Bergmann1982}. 
The anti-localization of non-stationary waves and the  weak anti-localization are different phenomena. The weak anti-localization
is related to the weak localization, whereas the anti-localization of non-stationary waves is
observed in systems where the strong localization can be possible \cite{Shishkina2023jsv}.

In the paper, we show that the anti-localization is not a specific
peculiarity of linear systems. We demonstrate that it is also common for nonlinear ones. 
To achieve this, we solve several
non-stationary problems for an infinitely long $\beta$-FPUT chain, introduced in the famous study \cite{Fermi1955},
with a defect.  
In the strongly nonlinear
case, we have no possibility to investigate the anti-localization in an analytic
way using the method of stationary phase. Accordingly, we use a pure numerical
approach, considering the previously obtained solutions for the corresponding
linear systems as the reference ones. All numerical calculations presented in the
paper have been performed twice using completely independent code and different numerical approaches. 
The first approach is to use routine 
{\tt scipy.integrate.odeint} included in
{\sc python} framework {\sc scipy} \cite{scipy}, which is actually based on {\sc lsoda} from the {\sc fortran} library {\sc odepack}.
The second approach is related to the self-written code utilizing the semi-implicit Euler method \cite{Skeel2020}. 
The results obtained by both approaches seem almost identical. The
choice of a discrete system (not a continuum one) as the system under investigation is caused by the considerations of
the simplicity of numerical solving.

In our opinion, the anti-localization essentially
influences the processes of heat transfer in linear and nonlinear lattices. In
particular, the anti-localization can explain the occurrence of the Kapitza thermal resistance 
\cite{Gavrilov2024,Gendelman2021}.
The emergence of cold points at the defects demonstrated in linear
and nonlinear cases \cite{Gendelman2021}, apparently, can be associated with this
phenomenon, though the mathematical formulation used in \cite{Gendelman2021} is different. 
The anti-localization can prevent the direct measurement of nonequilibrium
temperature of a lattice as discussed in \cite{Hatano2003}, since the attaching the
thermometer to the lattice leads to the distortion of the temperature field due to the
anti-localization.
The existence of the anti-localization makes a single-mode ansatz very
effective \cite{Gavrilov2024nody}
when looking for approximate non-stationary solutions describing
non-stationary oscillation in the systems where the strong localization occurs.

 The structure of the paper is as follows.
In Sect.~\ref{sec-formulation}, we discuss the mathematical formulation of the
problem in the dimensionless form.  In Sects.~\ref{sect-isotope}--\ref{sect-2springs},
several nonlinear problems, where we expect the emergence of the
anti-localization, are investigated. 
In Conclusion
(Sect.~\ref{sect-conclusion}), we discuss the basic results of the paper. 
In Appendix~\ref{AppA}, we provide the procedure of the non-dimensionalization
for the problem. In Appendix~\ref{App-B}, we reproduce some important formulas from
\cite{Shishkina2023cmat} to explain to the reader what the anti-localization of
non-stationary waves is in the linear case. In Appendix~\ref{AppC}, we provide
some reference formulas for the semi-infinite linear chain.

\section{The mathematical formulation}
\label{sec-formulation}

\begin{figure}[htbp]
  \centering\includegraphics[width=0.9\columnwidth]{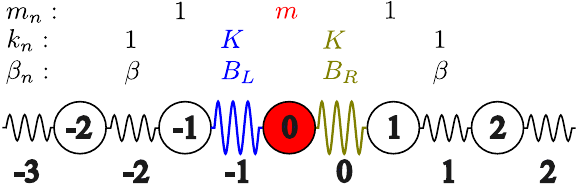}
\caption{The schematic of the system}
\label{crystal5.eps}
\end{figure}

Consider an infinite ordered chain of point particles with nearest-neighbor interactions. 
All particles have an equal mass, whereas  
one particle has an alternated mass. The particles are labeled by the discrete variable $n\in\mathbb Z$, where 
$\mathbb Z$ is
the set of all integers.
Without loss of generality, we assume that the alternated mass corresponds to $n=0$.
The particles are connected by nonlinear springs (i.e., by the bonds) with cubic nonlinearity also
labeled by $n\in\mathbb Z$. We
assume that a
spring label equals the label of the preceding particle.
All springs have equal stiffness excepting two of them with numbers $n=-1$ and $n=0$, which connect the particle
with $n=0$ and its neighbors. Thus, we consider a $\beta$-FPUT chain with a
point impurity (or a defect) in the form of a mass-spring inclusion. Everywhere
through the paper, excepting Sect.~\ref{sect-2springs}, we assume that the
inclusion is symmetric \footnote{In Sect.~\ref{sect-2springs} we, in
particular, consider a linear chain with an inclusion in the form of a
nonlinear spring with the linear stiffness being equal to the spring stiffness in the
rest of the chain.}. 
The schematic of the system is presented in
Fig.~\ref{crystal5.eps}.

The equations of motion in the dimensional form are formulated in
Appendix~\ref{AppA}, where the procedure on non-dimensionalization is
presented. 
In the dimensionless form, we have the following 
infinite system of differential-difference equations:
%
\begin{gather} 
  \dot u_n=v_n,
  \label{baseeq1}
\\
\begin{multlined} 
m_n\dot v_n=
k_{n}(u_{n+1}-u_n)-k_{n-1}(u_{n}-u_{n-1})\\
+\beta_{n}(u_{n+1}-u_n)^3-\beta_{n-1}(u_n-u_{n-1})^3,
  \label{baseeq2}
\end{multlined} 
\end{gather} 
where
\begin{equation}
  \begin{aligned} 
  &m_n=1+(m-1)\delta_n,
  \\
  &k_n=1+(K-1)(\delta_{n+1}+\delta_n),
  \\
  &\beta_n=\beta+(B_L-\beta)\delta_{n+1}+(B_R-\beta)\delta_n.
  \label{eq:constants}
  \end{aligned} 
\end{equation}
Here,
overdot denotes the derivative with respect to
the dimensionless time $t$;
$u_n(t)$ and $v_n(t)$ are the dimensionless displacement and particle velocity,
respectively, of the particle with a label
$n\in\mathbb Z$; $\beta$ is the dimensionless cubic stiffness for the bonds outside
the defect; $m>0$ is the dimensionless mass of the particle at the defect;
$K$ and $B$ are the linear and cubic dimensionless stiffnesses, respectively, for the bonds at the
defect;
$\de_n$ is {the Kronecker delta} ($1$ if and only if $n=0$, $0$ otherwise, {$n\in\mathbb Z$}).

The initial conditions are:
\begin{equation} 
  u_n(0)=0, \qquad v_n(0)=m_N^{-1}\delta_{n-N}.
\label{ic}
\end{equation} 
Initial conditions in the form of Eq.~\eqref{ic} correspond to the pulse
loading $\delta(t)$ applied to the particle with label $N$.
Here $\delta(t)$ is the Dirac delta-function.

Note that we consider the strongly nonlinear system, i.e., the dimensionless
parameter 
\begin{equation}
\bar\beta=\max_n\big(|\beta_n|\big)
\end{equation}
is assumed to be of order one, i.e., not a small quantity. Thus, we expect 
that the clear difference between a nonlinear solution and the corresponding
linear one can be observed for finite values of time $t$.

In the particular case 
\begin{equation}
 m=1,\qquad K=1,\qquad B_L=B_R=\beta=0
\end{equation}
of an infinite uniform linear chain, the exact solution of the problem defined by Eqs.~\eqref{baseeq1}--\eqref{ic} 
for the particle velocities $v_n(t)$
is $V_{n-N}$ \cite{schrodinger1914dynamik,Hamilton1940,Havelock1910,Muehlich2020}:
\begin{equation}
V_{n}=J_{2n}(2t)=J_{2|n|}(2t),
\label{Sro-bessel}
\end{equation}
where $J_n(\cdot)$ is the Bessel function of the first kind of order
$n\in\mathbb Z$.


\section{An infinite $\beta$-FPUT chain with an isotopic defect}
\label{sect-isotope}

We assume 
\begin{equation}
  B=B_L=B_R
  \label{eq:B-def}
\end{equation}
unless the opposite is explicitly indicated.
Consider a $\beta$-FPUT chain with an isotopic impurity (a defect)
\begin{equation}
  m\neq1,\quad m>0,
  \label{m-def}
\end{equation}
in chain where
\begin{gather}
  B=\beta,
  \label{eq:B}
  \\
  K=1.
\end{gather}
The defect is subjected to a pulse loading at position 
\begin{equation}
N=0.
  \label{eq:N=0}
\end{equation}

In the linear case $\beta=0$, this problem is considered in studies
\cite{Rubin1963,Shishkina2023cmat}.
In \cite{Shishkina2023cmat}
an approximate solution of asymptotic nature,
describing the wave-field as a whole
and valid for large values of time $t$,
is obtained.
We reproduce in Appendix~\ref{App-B} some important formulas from \cite{Shishkina2023cmat} to
explain to the reader what the anti-localization of non-stationary linear waves is.

Now let us consider how a nonlinearity influences the anti-localization.
To do this, following to  \cite{Shishkina2023cmat}, we proceed with numerical integration of $2n_0+1$ ODE
\eqref{baseeq1}, \eqref{baseeq2} {with $n\in -n_0\dots n_0$, $n_0>\lambda t$},
$\lambda=\const>1$.
At first, consider the case of a heavy defect, where the impurity-induced localized mode does not exist 
for the corresponding linear system with $\beta_n=0$, and, hence,
$v_n^{\mathrm{stop}}=0$. 

\begin{figure}[htbp]
\centering\includegraphics[width=1\columnwidth]{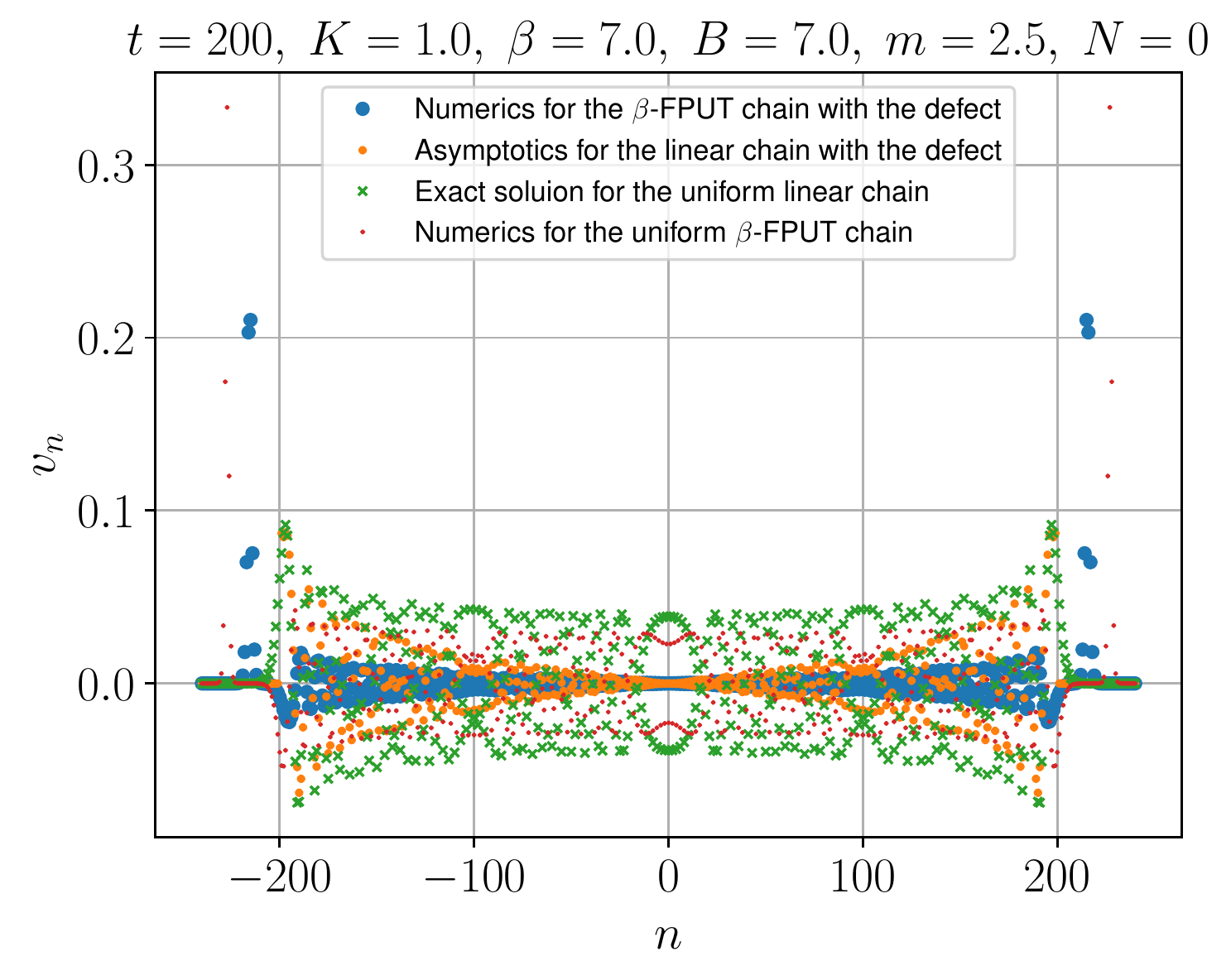}
\caption{The particle velocity $v_n(t)$ versus the spatial variable $n$ in an infinite $\beta$-FPUT chain
  and the corresponding linear chain with and without a heavy isotopic defect.
The anti-localization, i.e., zeroing of the perturbation amplitude near $n=0$, is clearly observed for both 
linear and nonlinear systems with the defect and not observed for systems
without the defect. The difference between nonlinear and linear solutions for the system with the defect is
  clearly recognizable everywhere excepting the anti-localization zone.}
\label{V_iso_heavy.pdf}
\end{figure}

In Fig.~\ref{V_iso_heavy.pdf}, the numerically found
particle velocity $v_n(t)$ versus the spatial variable $n\in\mathbb Z$ is shown for a fixed value of time.
For reference, the wave-field in the corresponding linear system with $\beta_n=0$ given by
Eqs.~\eqref{dot-u-fourier-gen}, \eqref{c-trapped}--\eqref{psi}, 
the wave-field in the
uniform linear chain given by Eq.~\eqref{Sro-bessel}, 
and the numerically found wave-field in the corresponding uniform $\beta$-FPUT chain 
are also displayed.
One can compare Fig.~\ref{V_iso_heavy.pdf} with Fig.~2a in
\cite{Shishkina2023cmat}, where it is demonstrated that the numeric results for the linear system 
are in excellent agreement with asymptotic description 
Eqs.~\eqref{dot-u-fourier-gen}, \eqref{c-trapped}--\eqref{psi}, 
everywhere excepting the leading wave-fronts $n\simeq t$. The nonlinearity
essentially affects the wave-fields 
near the leading fronts $n\simeq t$, see \cite{Li2010}. 
In a central zone far from the leading fronts, we observe that the nonlinearity quantitatively 
alters the solution: the amplitude of oscillation in the nonlinear case is
approximately a half of the amplitude in the linear case.
Zeroing of the oscillation amplitude near $n=0$,
i.e., the anti-localization, is observed in both linear and nonlinear
systems with the defect. On the other hand, the
anti-localization is not observed in both uniform systems under investigation.

\begin{figure}[htbp]
\centering\includegraphics[width=1\columnwidth]{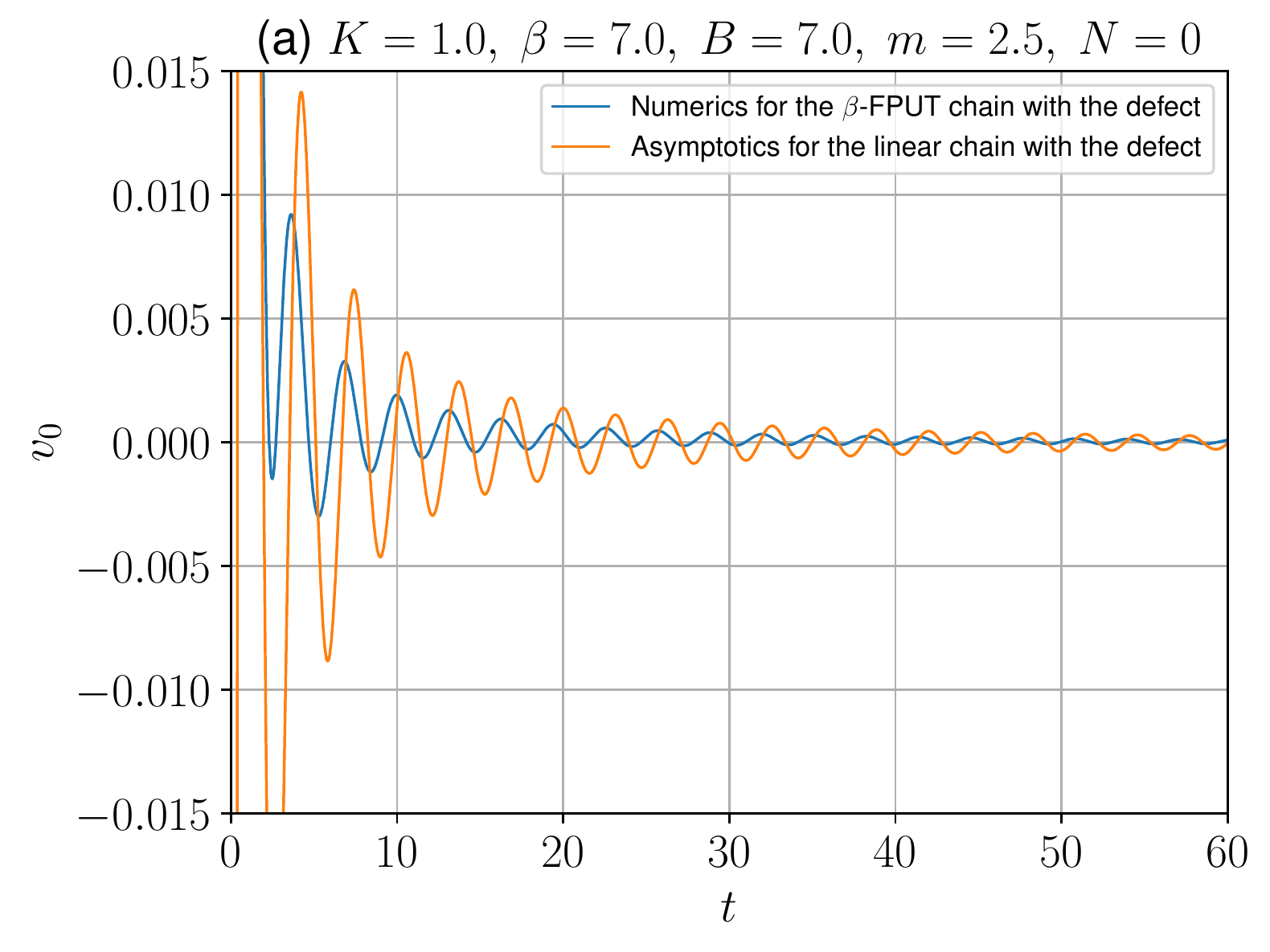}
\centering\includegraphics[width=1\columnwidth]{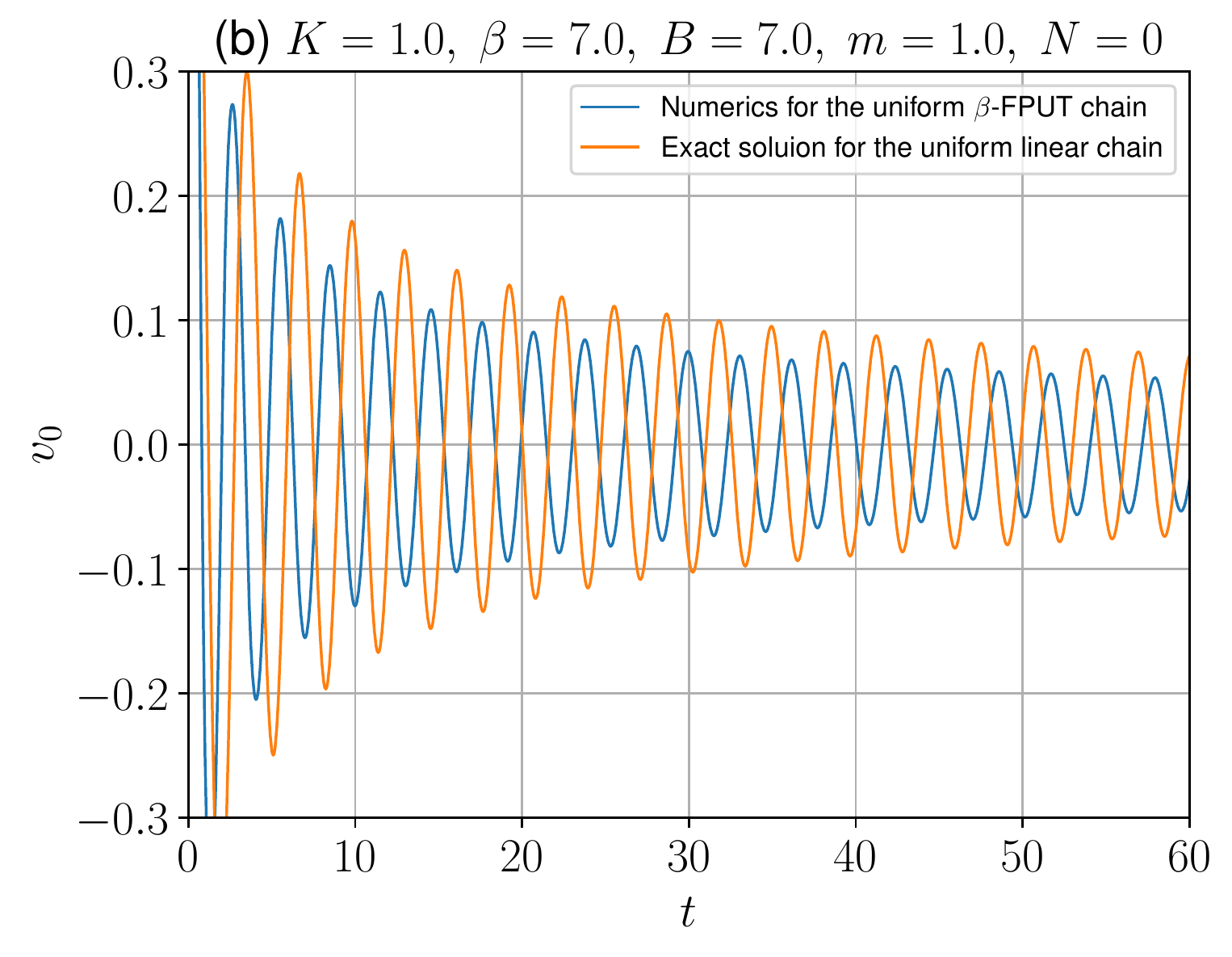}
\caption{The particle velocity $v_0(t)$ versus time in an infinite $\beta$-FPUT chain and in the corresponding linear chain. 
(a) The solutions for a system with a heavy isotopic defect; (b) the solutions
for the corresponding uniform system. In both sub-plots, the order of decay for the
nonlinear system can be
estimated as the same as it is observed in the corresponding linear system.
Oscillation asymmetry is clearly recognizable for the nonlinear system with
the defect.}
\label{V_iso_heavy_t.pdf}
\end{figure}

In Fig.~\ref{V_iso_heavy_t.pdf}a the numerically found solution for 
the particle velocity $v_0(t)$ versus the time $t$ is shown for the
nonlinear system with the heavy defect.
For reference, the same particle velocity in the corresponding linear system with $\beta_n=0$,
i.e., the asymptotics given by 
Eq.~\eqref{rubin-f} and the numerical solution, are also displayed. 
One can see that in the linear system the numerical solution very often converges to the asymptotics.
On the other hand, we clearly deal with an essentially nonlinear regime,
since the solutions in linear and nonlinear cases are not close to each other
for finite times.
According to Fig.~\ref{V_iso_heavy_t.pdf}a,
the order of decay for oscillation in the nonlinear system can be estimated  
the same 
as it is observed in the corresponding linear system, i.e., $t^{-3/2}$, see Eq.~\eqref{rubin-f}.
The essential peculiarity of the oscillation in the nonlinear
system with the defect, which can be observed in Fig.~\ref{V_iso_heavy_t.pdf},
is the oscillation asymmetry. Indeed, one can see that there exists a vanishing
(apparently with the same power law $t^{-3/2}$) 
non-negative component of the particle velocity $v_0(t)$. Such a
component does not exist in the corresponding linear system \cite{Shishkina2023cmat}.

In Fig.~\ref{V_iso_heavy_t.pdf}b the numerically found
solution for the particle velocity $v_0(t)$ versus the time $t$ is shown for the
nonlinear system without the defect.
For reference, the same particle velocity in the corresponding linear system with $\beta_n=0$ given by 
Eq.~\eqref{Sro-bessel},
is also displayed.
The order of 
decay for oscillation in the nonlinear system can be estimated as $t^{-1/2}$
as it is observed according to Eqs.~\eqref{sum-I_1+I_2-single-wave-amp},
\eqref{phi_ast-expr}, \eqref{A(0)}
in the uniform linear system. 
Thus, there is no anti-localization in both nonlinear and linear systems
without a defect.

\begin{figure}[htbp]
\centering\includegraphics[width=1\columnwidth]{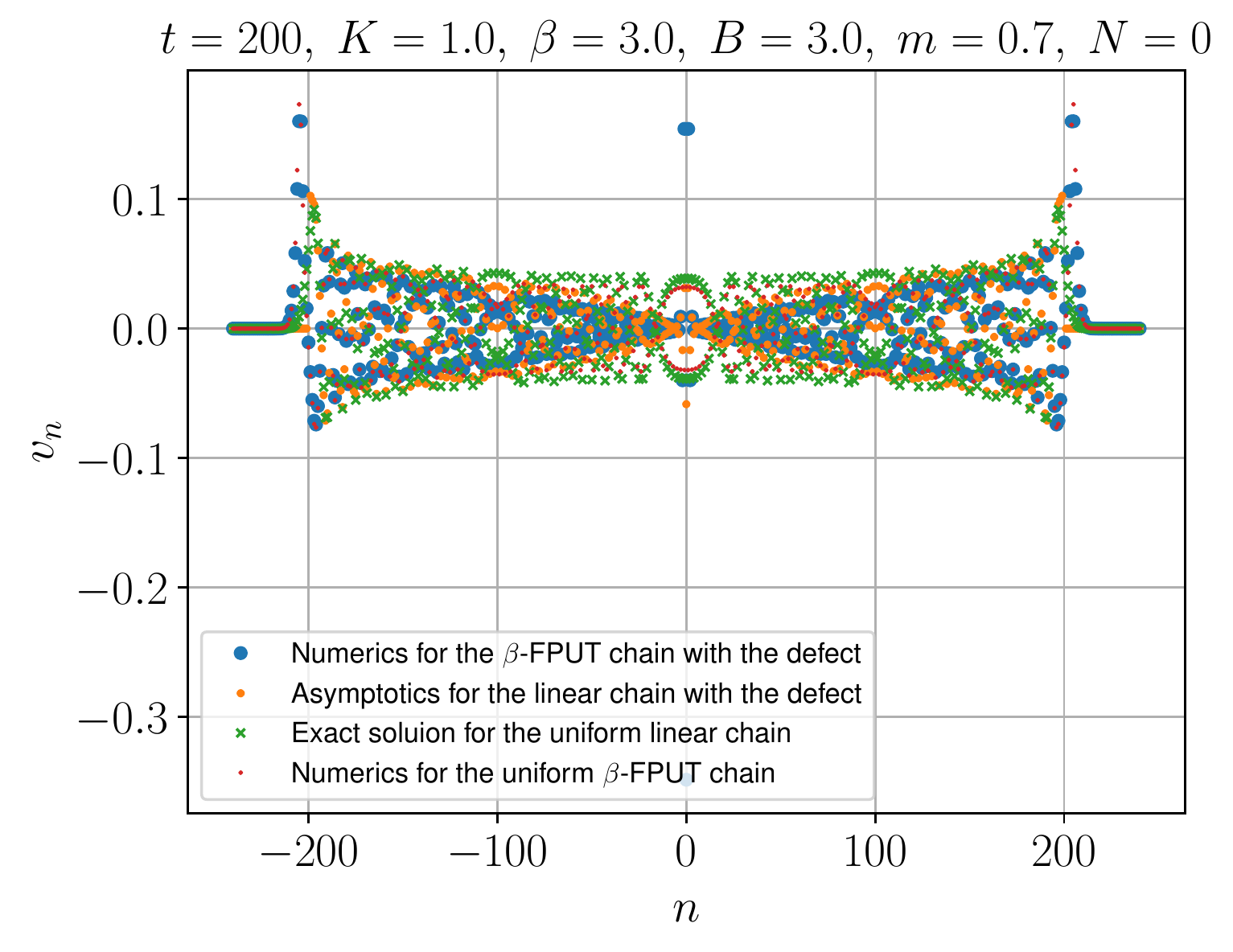}
\caption{The particle velocity $v_n(t)$ versus the spatial variable $n$ in an infinite $\beta$-FPUT chain
  and the corresponding linear chain with and without a light isotopic defect. The anti-localization in systems with
the defect co-exists with impurity-induced localized mode. The latter forms a peak of the amplitude 
near $n=0$}
\label{V_iso_light.pdf}
\end{figure}

\begin{figure}[hbtp]
\centering\includegraphics[width=1\columnwidth]{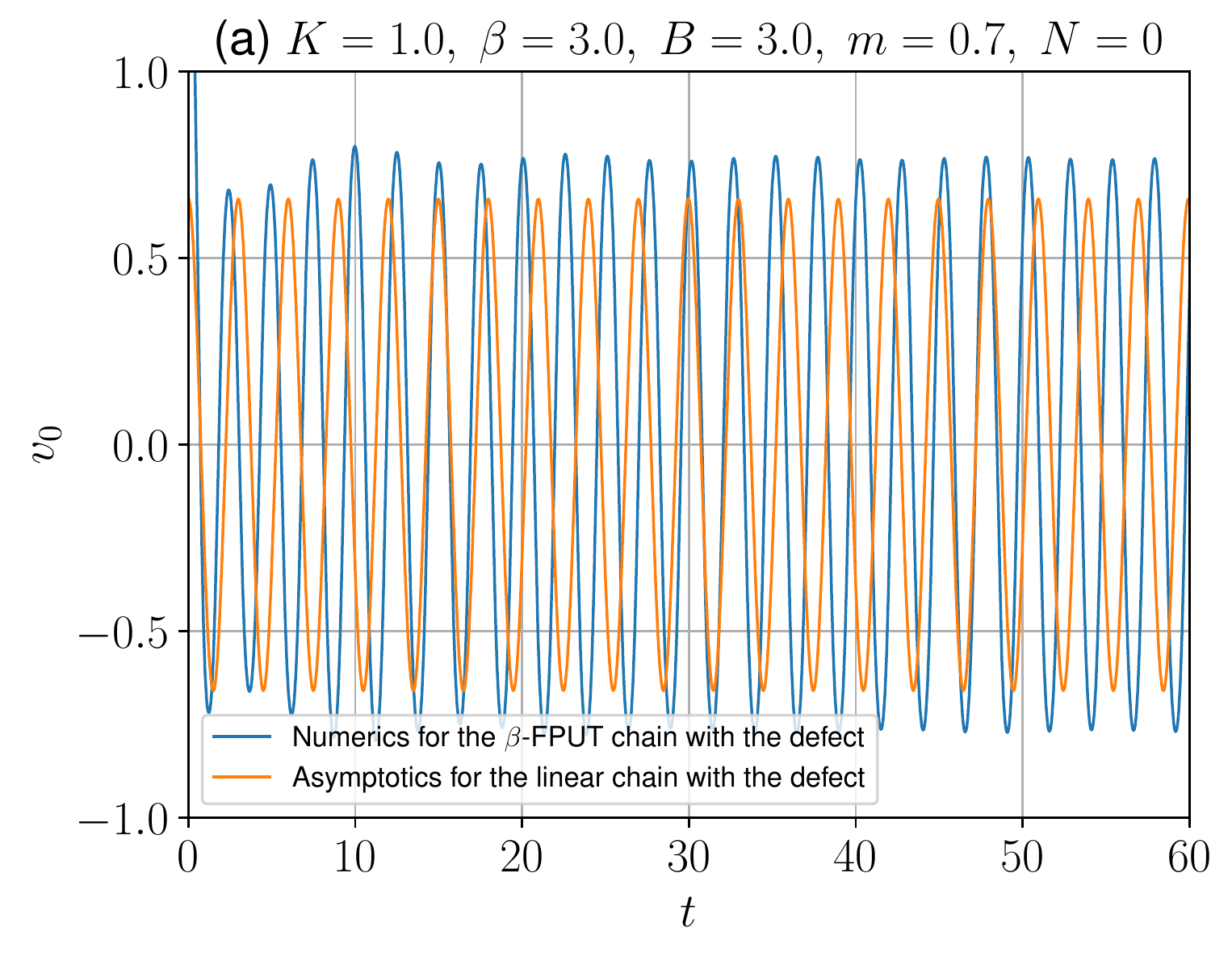}  
\caption{The particle velocity $v_0(t)$ versus time in an infinite $\beta$-FPUT chain and in the corresponding linear chain 
in the case of a light isotopic defect. For both linear and nonlinear systems,
a non-vanishing localized oscillation is observed. The frequency shift
observed for the nonlinear system confirms the nonlinear character of the oscillation}
\label{V_iso_light_t.pdf}
\end{figure}

Finally, consider the case of a light defect $0<m<1$. 
As it is shown in \cite{Shishkina2023cmat},
in the corresponding linear system, the localized component $\VA_n^{\mathrm{stop}}$
co-exists with the anti-localized one $\VA_n^{\mathrm{pass}}$. 
We also know that adding nonlinearity does not destroy an impurity-induced
localized mode in the case of the chain \cite{Kivshar1991}.
In Fig.~\ref{V_iso_light.pdf},
where the particle velocity $v_n(t)$ versus the spatial variable $n\in\mathbb Z$ is shown for a fixed value of time,
one can see that in the nonlinear system we also observe a similar wave pattern,
which corresponds to the superposition of  localized and anti-localized components.
In Fig.~\ref{V_iso_light_t.pdf}, the numerically found solution for 
the particle velocity $v_0(t)$ versus the time $t$ is shown for the
nonlinear system with the defect.
For reference, the same particle velocity in the corresponding linear system with $\beta_n=0$ given by 
Eq.~\eqref{rubin-f} is also displayed. In both cases (the nonlinear and the linear ones), a non-vanishing impurity 
induced localized oscillation is observed. The frequency shift
observed for the nonlinear system confirms the nonlinear character of the localized oscillation.

Qualitatively similar results can be exhibited for various values of the system
parameters, such as for a hard $\beta$-FPUT chain with $\beta>0$ as well as for a soft one with $\beta<0$.


\section{An infinite $\beta$-FPUT chain with a mass-spring inclusion}
\label{sect-osc}
\begin{figure}[htbp]
\centering\includegraphics[width=1\columnwidth]{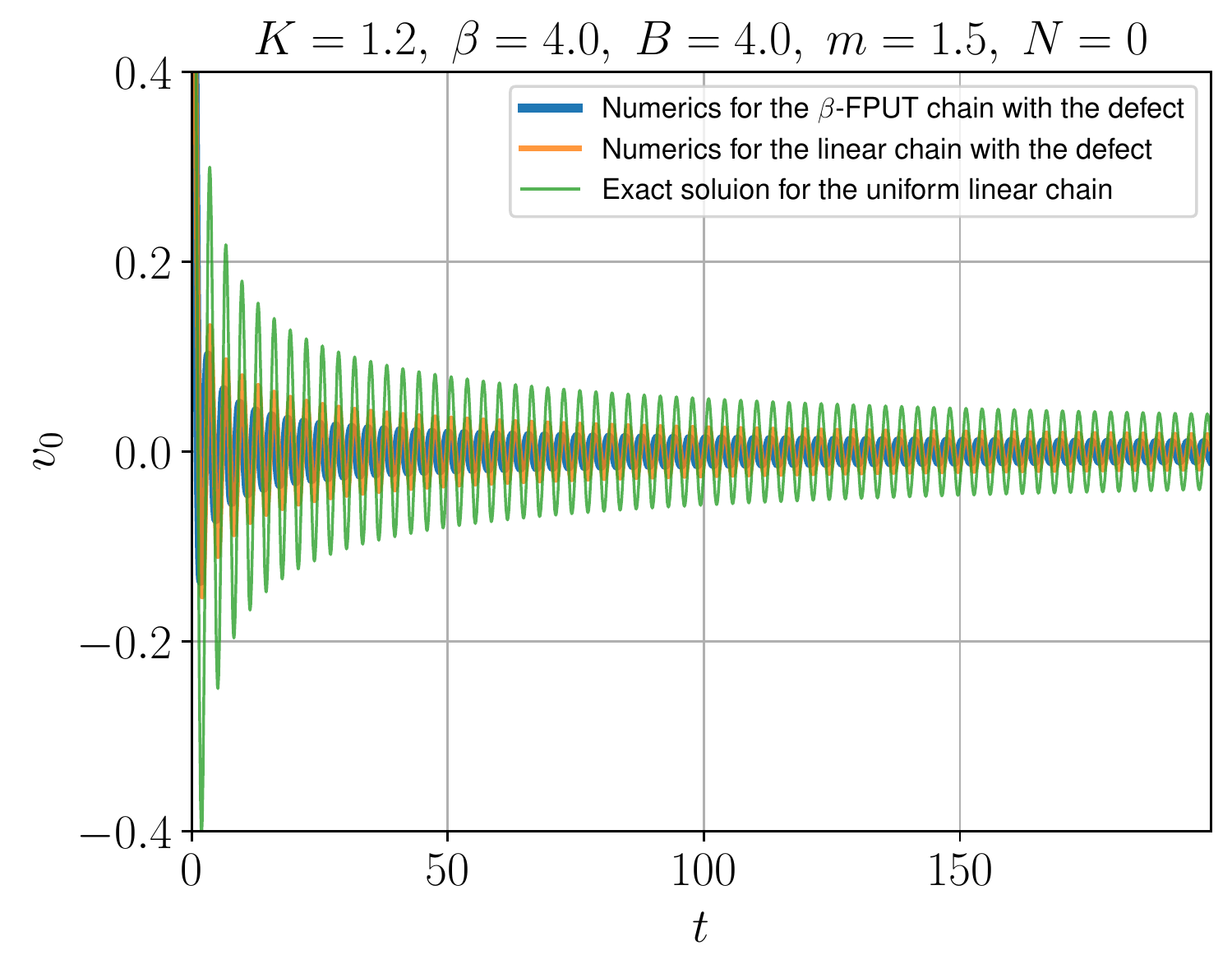}
\caption{The particle velocity $v_0(t)$ versus time in the case of the mass-spring
defect. Parameters of defect satisfy equation $K=K_0$. The order of decay for all three systems 
can be estimated as the same}
\label{V_boundary_t.pdf}
\end{figure}

\begin{figure}[htbp]
\centering\includegraphics[width=1\columnwidth]{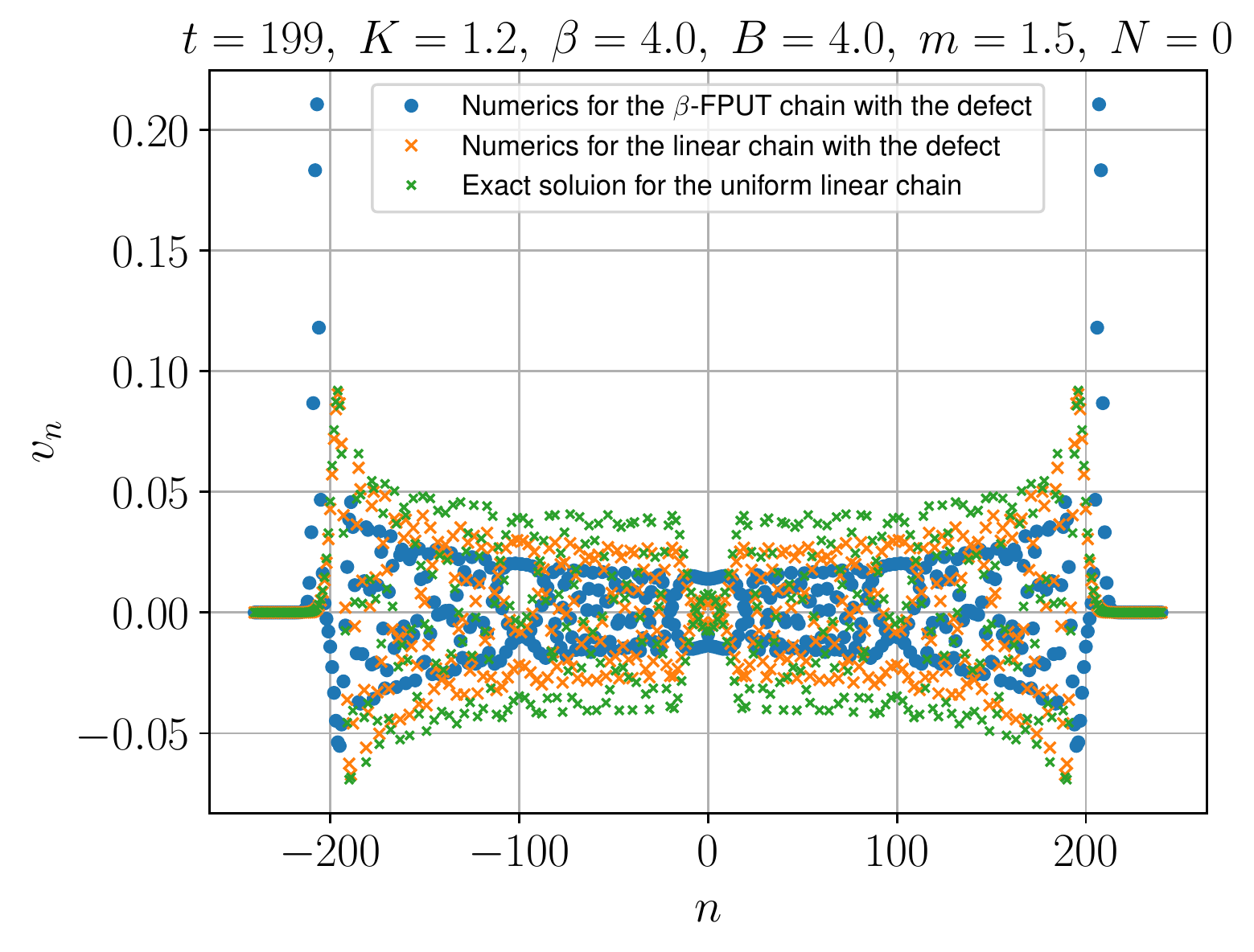}
\caption{The particle velocity $v_n(t)$ versus the spatial variable $n$ 
in the case of the mass-spring
defect. Parameters of defect satisfy equation $K=K_0$. The anti-localization
is absent in all three cases (the minimum at $n=0$ observed for the linear
systems is not an amplitude minimum and related to the value of the phase $\phi t$)
}
\label{V_boundary.pdf}
\end{figure}

\begin{figure}[htbp]
\centering\includegraphics[width=1\columnwidth]{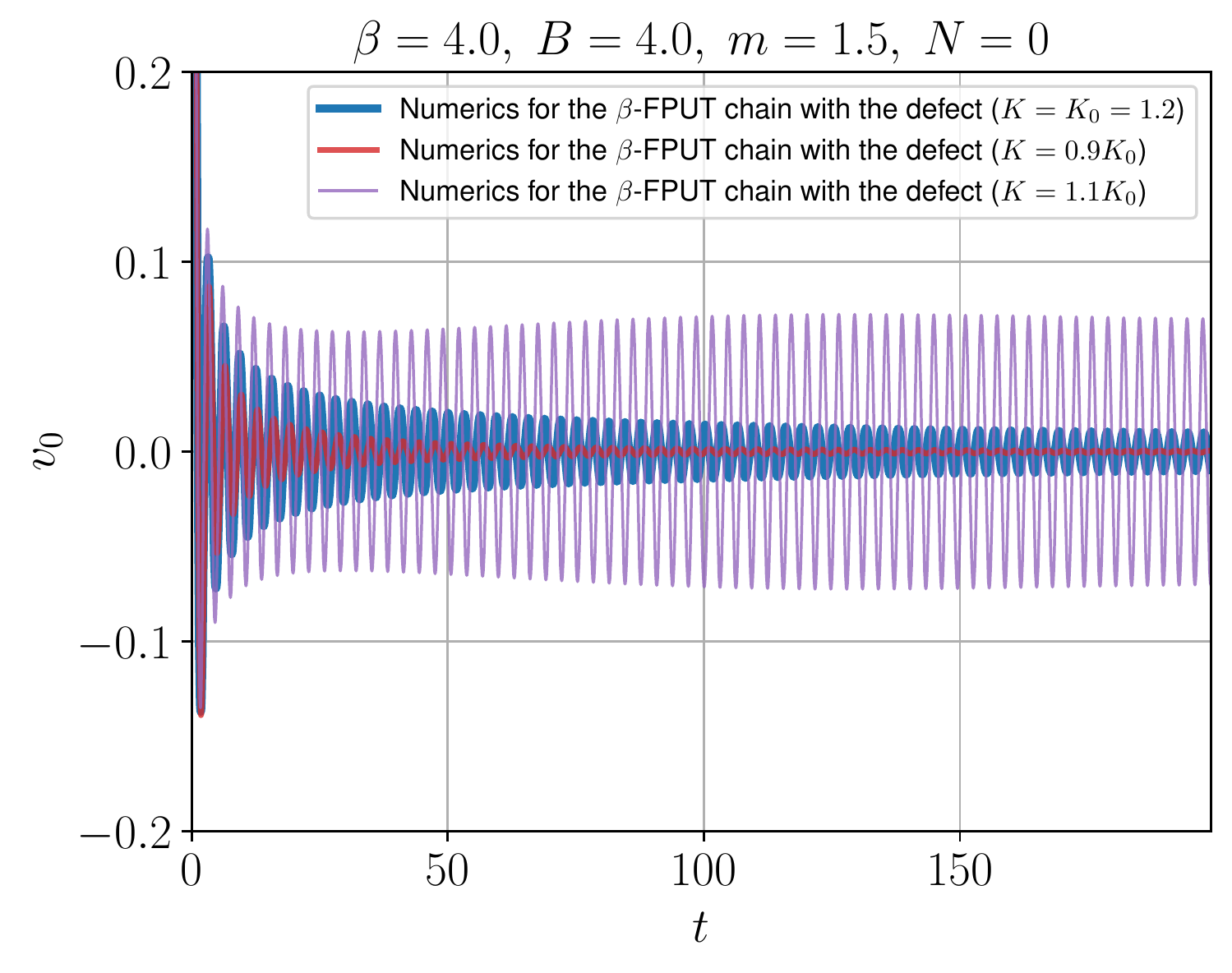}
\caption{The particle velocity $v_0(t)$ versus time in an infinite $\beta$-FPUT chain with
a mass-spring defect. The orders of decay are clearly different in all three cases.}
\label{V_boundary_t1.pdf}
\end{figure}

As we can see from the previous Sect.~\ref{sect-isotope}, the anti-localization
exists in linear and nonlinear systems with an isotopic defect. The anti-localization is
observed for both cases $m>1$ and $m<1$, though, in the latter case, the
anti-localization co-exists with the localization. The boundary in the system
parameter space between these two cases corresponds to the uniform system,
where there is no anti-localization. 
In previous
study \cite{Shishkina2023jsv},
we have shown that besides the uniform system, the anti-localization
disappears in linear non-uniform systems with parameters, which correspond to such a boundary separating domains in
the system parameter space where the
localization is observed and not observed. The linear uniform system
is just a particular case of
such a system with the boundary parameter values. In this section, we want
to check if the anti-localization disappears for non-uniform {\it nonlinear} systems with
such boundary values of their parameters. To achieve this, we consider the chain
with parameters satisfying Eq.~\eqref{eq:B}, equipped with a mass-spring inclusion for 
which Eq.~\eqref{m-def} and 
\begin{equation}
K\neq1
\end{equation}
are fulfilled.
The defect is subjected to a pulse loading at the position defined by  Eq.~\eqref{eq:N=0}.
In the corresponding linear system with $B=\beta=0$ the impurity-induced
localization is observed if and only if the following inequality is
fulfilled \cite{Montroll1955,Yu2019,Shishkina2023asm}:
\begin{gather}
 K>K_0,
 \label{KK}
 \\
 K_0\=\frac{2m}{1+m}.\label{K0-def}
\end{gather}
Inequality \eqref{KK} defines a domain in the problem parameter space, which we call
the localization domain. Now consider a nonlinear system with a fixed value $B=\beta$,
and fixed $m$ as we have taken in Sect.~\ref{sect-isotope}. 
Now we try to determine the boundary value of the parameter $K$, which separates nonlinear 
systems where the impurity-induced
localization is observed or not observed. The numeric experiment shows that
such a value is very close to the value $K=K_0$ 
defined by Eq.~\eqref{K0-def} for the linear systems, see Figs.~\ref{V_boundary_t.pdf}--\ref{V_boundary_t1.pdf}.

In Fig.~\ref{V_boundary_t.pdf} the numerical solution $v_0(t)$ versus the time $t$ for $\beta$-FPUT chain with
a mass-spring defect is presented. 
The value of the linear spring stiffness is taken  
equal to the boundary value $K_0$. 
For reference, the same particle velocity in the corresponding linear system with the same defect,
as well as the exact solution
\eqref{Sro-bessel}
for the uniform linear chain, are also presented \footnote{Unfortunately, we
do not have here the reference asymptotic non-stationary solution for the
corresponding linear system.}.
The order of 
decay for oscillation in both nonlinear systems with and without the defect can be estimated as $1/\sqrt t$
as it is observed according to 
Eqs.~\eqref{sum-I_1+I_2-single-wave}, \eqref{A(0)}
in the uniform linear chain. 
In Fig.~\ref{V_boundary.pdf}, the
particle velocities $v_n(t)$ versus the spatial variable $n\in\mathbb Z$ are 
shown at a fixed value of time for the same three systems as in Fig.~\ref{V_boundary_t.pdf}. 
The absence of the anti-localization in the system with the defect for which
$K=K_0$ is clearly recognizable.

In Fig.~\ref{V_boundary_t1.pdf} the same numerical solution $v_0(t)$   for $\beta$-FPUT chain with
a mass-spring defect with $K=K_0$, which is already shown in Fig.~\ref{V_boundary_t.pdf}, is compared 
with analogous solutions with $K=0.9K_0$ and $K=1.1K_0$. We can see that the
solution corresponding to $K=0.9K_0$ vanishes faster than one taken for $K=K_0$. The corresponding order of decay
is expected to be $1/\sqrt{t^3}$, see Sect.~\ref{sect-isotope}. On the other hand, the
solution corresponding to $K=1.1K_0$ is a non-vanishing one, i.e., an impurity
induced localized oscillation is observed.

Thus, we have shown that the
anti-localization is not observed in both linear and nonlinear systems with a mass-spring defect if 
the system parameters correspond to 
the boundary of the localization domain in the problem parameter space. On
the other hand, the anti-localization is observed if the system parameters lie
outside this boundary.




\section{A semi-infinite $\beta$-FPUT chain}
\label{sect-semi}
Assuming that $K=0$, $B=0$, we split an infinite chain into two uncoupled semi-infinite
chains. Thus, a semi-infinite chain loaded at the end $n=N=1$ is a particular
case of a chain with a defect.

For the corresponding linear semi-infinite system, the effect that we call the anti-localization was first time discovered, 
as far as we know \footnote{We are grateful to
S.D.~Liazhkov.}, much earlier in 
\cite{Jackson1978} and explained by
specific peculiarities of linear systems. The basic formulas describing
anti-localized wave-field in the linear case the reader can find in Appendix~\ref{AppC}.
Let us check if the anti-localization exists in a semi-infinite $\beta$-FPUT chain. 

\begin{figure}[htbp]
\centering\includegraphics[width=1\columnwidth]{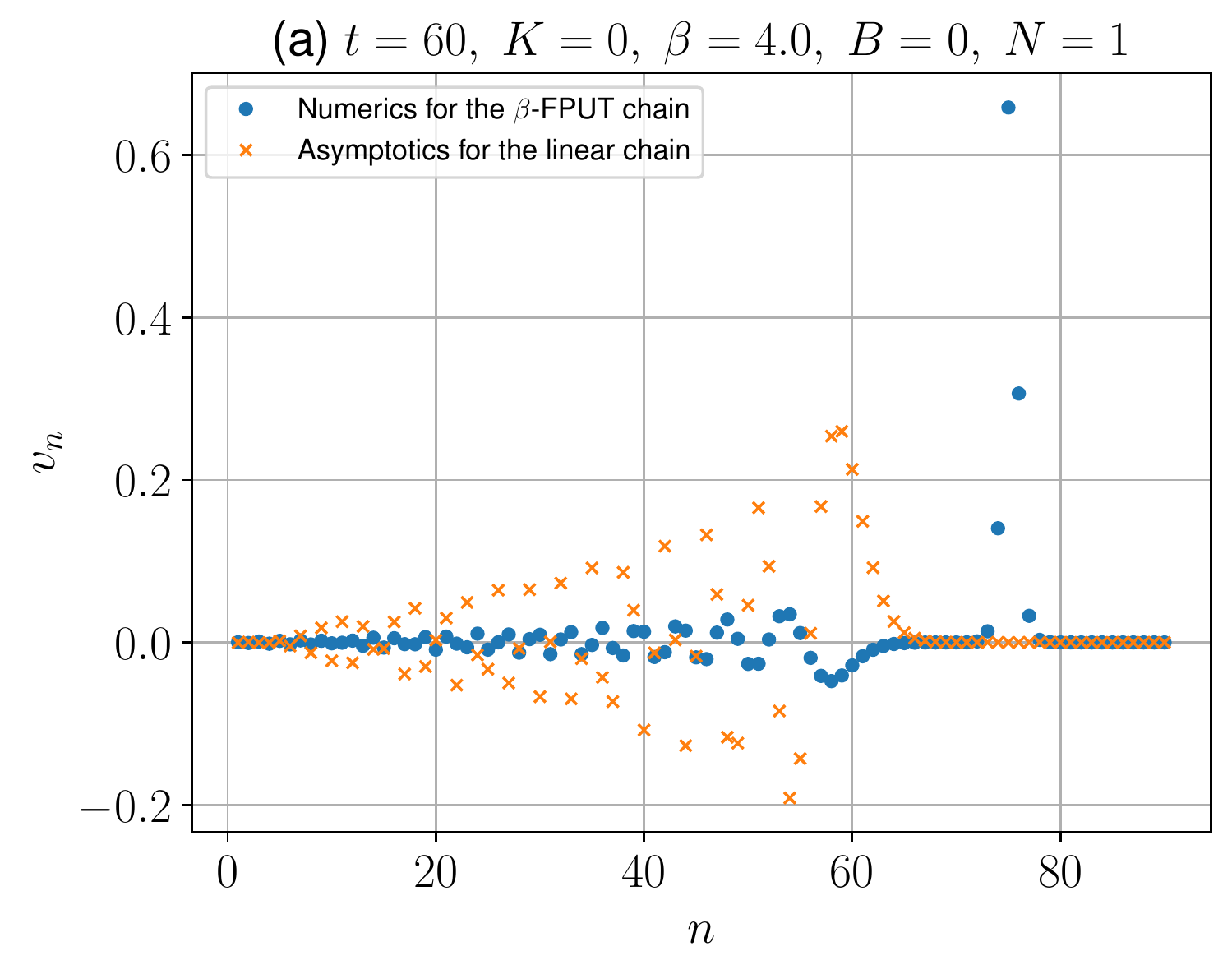}
\centering\includegraphics[width=1\columnwidth]{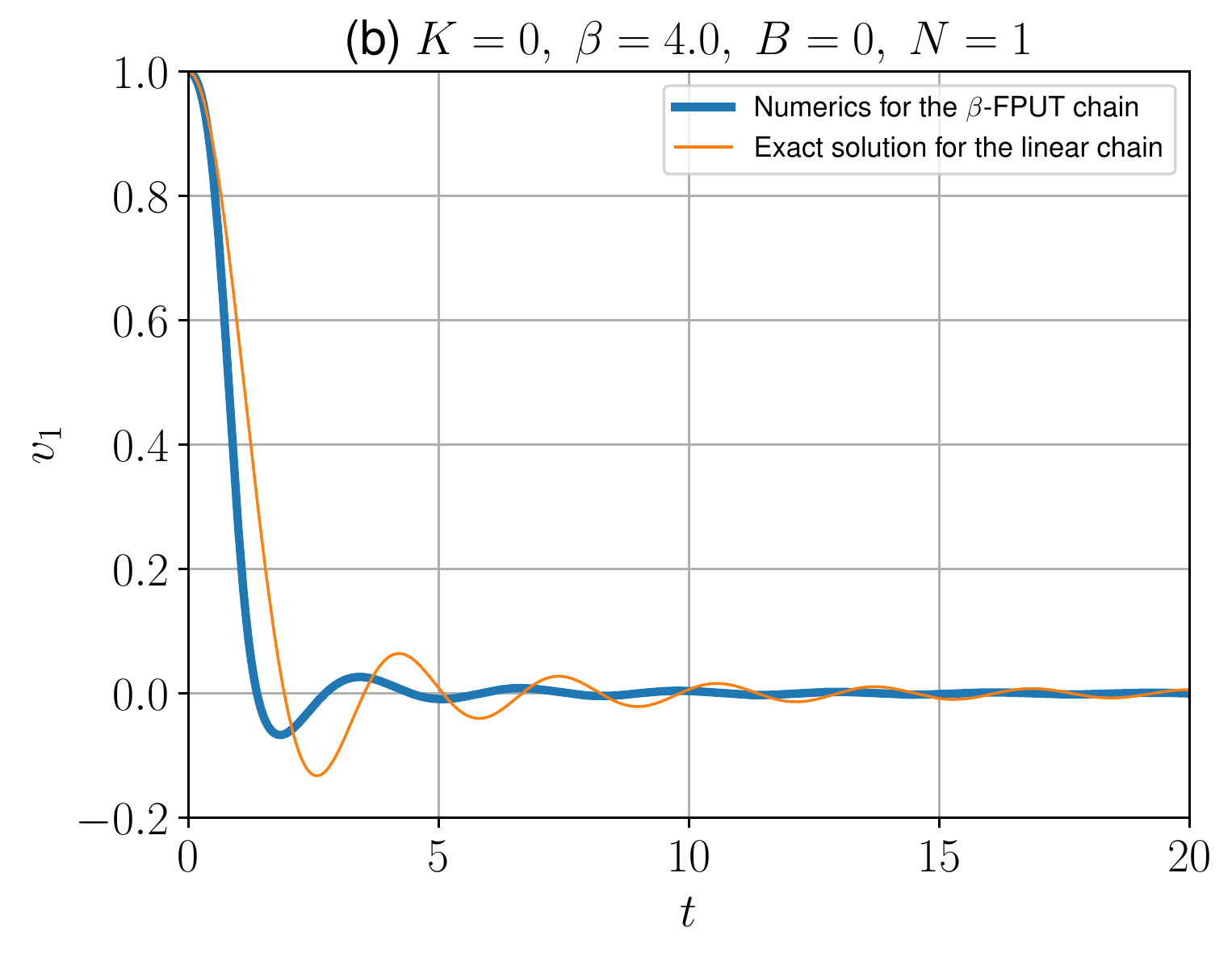}
\caption{The wave-field in the $\beta$-FPUT and linear chains loaded at the
  free end: (a) particle velocities $v_n$ versus the spatial variable $n$;
(b) particle velocity of the end particle  $v_1$ versus time. The
anti-localization is clearly observed}
\label{V_semi.pdf}
\end{figure}

The numerical solution, which corresponds to a nonlinear chain with $\beta\neq0$ is presented
in Fig.~\ref{V_semi.pdf}. 
For reference, the wave-field in the corresponding linear semi-infinite chain 
given by the 
exact solution~\eqref{semi-exact}
is also displayed. In Fig.~\ref{V_semi.pdf}a the particle velocities $v_n$ versus the spatial variable $n$ 
are presented. In Fig.~\ref{V_semi.pdf}b the particle velocity of the end particle $v_1(t)$ versus time is shown.
One can see that the anti-localization is clearly visible for both linear and
nonlinear systems.

\begin{remark} 
In Fig.~\ref{V_semi.pdf}a one can also observe a soliton-like wave propagating faster than the leading wave-front 
in the linear system. The solitons in a $\beta$-FPUT chain are discussed, e.g.,
in \cite{Vainchtein2022}.
\end{remark} 


\section{A nonlinear spring in an infinite linear chain}
\label{sect-2springs}

In this section, we want to show that introducing the local nonlinearity into
a linear system can also lead to a phenomenon similar to the anti-localization. We consider a uniform
linear chain, i.e., 
\begin{equation}
 m=1,\quad K=1,\quad \beta=0.
\end{equation}

\begin{figure}[p]
\centering\includegraphics[width=1\columnwidth]{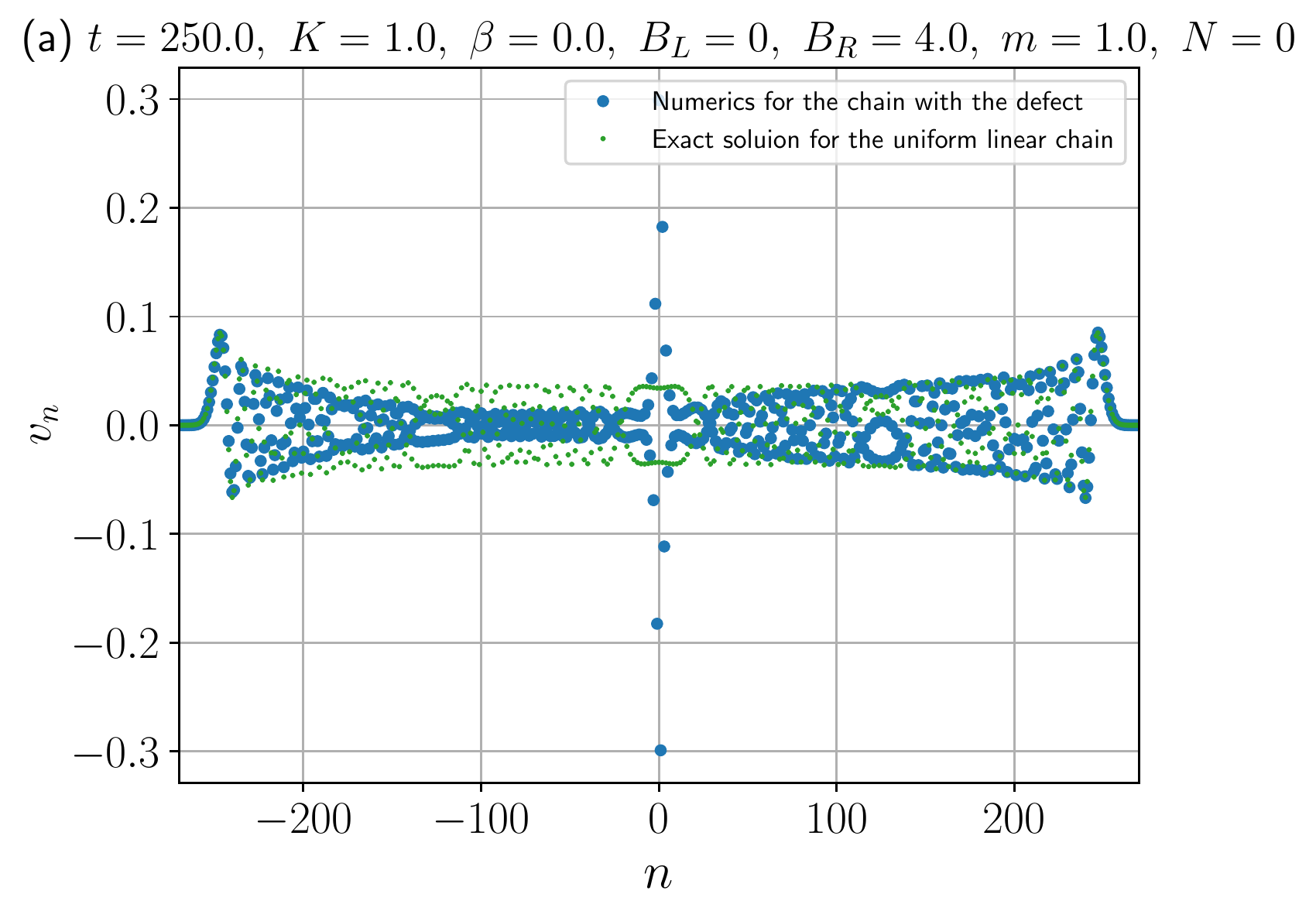}
\centering\includegraphics[width=1\columnwidth]{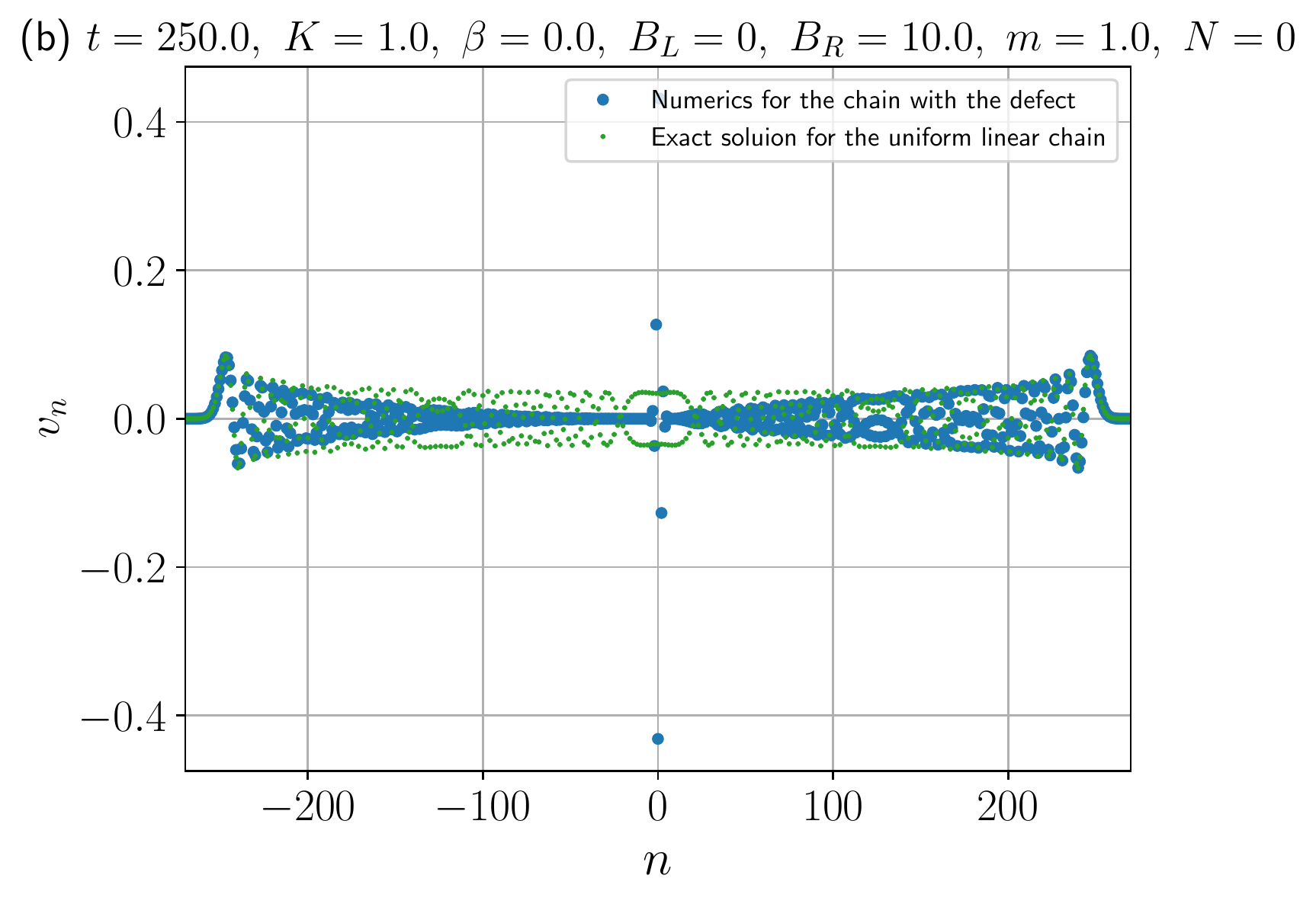}
\caption{The particle velocity $v_n$ versus the spatial variable $n$ in an infinite linear 
chain with and without a nonlinear spring. The greater $B_R$, 
the tendency to the 
anti-localization is more clear}
\label{V_linear.pdf}
\end{figure}

At first, consider the source at $N=0$ at the inclusion in the form of a hard nonlinear spring:
\begin{equation}
 B_L=0,\quad B_R>0.
\end{equation}
In Fig.~\ref{V_linear.pdf} we present the spatial distribution of the particle velocities
in the corresponding numeric experiment with pulse loading applied at the
particle with number $n=N=0$ just to the left from 
the nonlinear spring.
Sub-plots (a) \& (b) correspond to the choices $B_R=4.0$ and $B_R=10.0$,
respectively.
For reference, the particle velocities in the corresponding linear system with $B_L=B_R=0$ given by 
Eq.~\eqref{Sro-bessel} are also displayed in each sub-plot.
One can see that the wave pattern observed in Fig.~\ref{V_linear.pdf}b
is a bit similar to the one
presented in Fig.~\ref{V_iso_light.pdf}, where the anti-localization co-exists
with the impurity-induced localized oscillation. The clear difference is an expected asymmetry of 
the wave patterns in Fig.~\ref{V_linear.pdf}. In  Fig.~\ref{V_linear.pdf}a we
also observe weakening of the propagating component near the defect but not
zeroing. Increasing the coefficient $B_R$ one can see that the amplitude of
the propagating component near the defect decreases. 
Thus, we observe a weakening of the propagating component but, apparently, not an {\it
asymptotic} one. 
The observed phenomenon can be called {\it the
incomplete anti-localization}.
Here we cannot estimate, even qualitatively,
the order of decay for the anti-localized component due to the presence of the
localized mode.

\begin{figure}[htbp]
\centering\includegraphics[width=1\columnwidth]{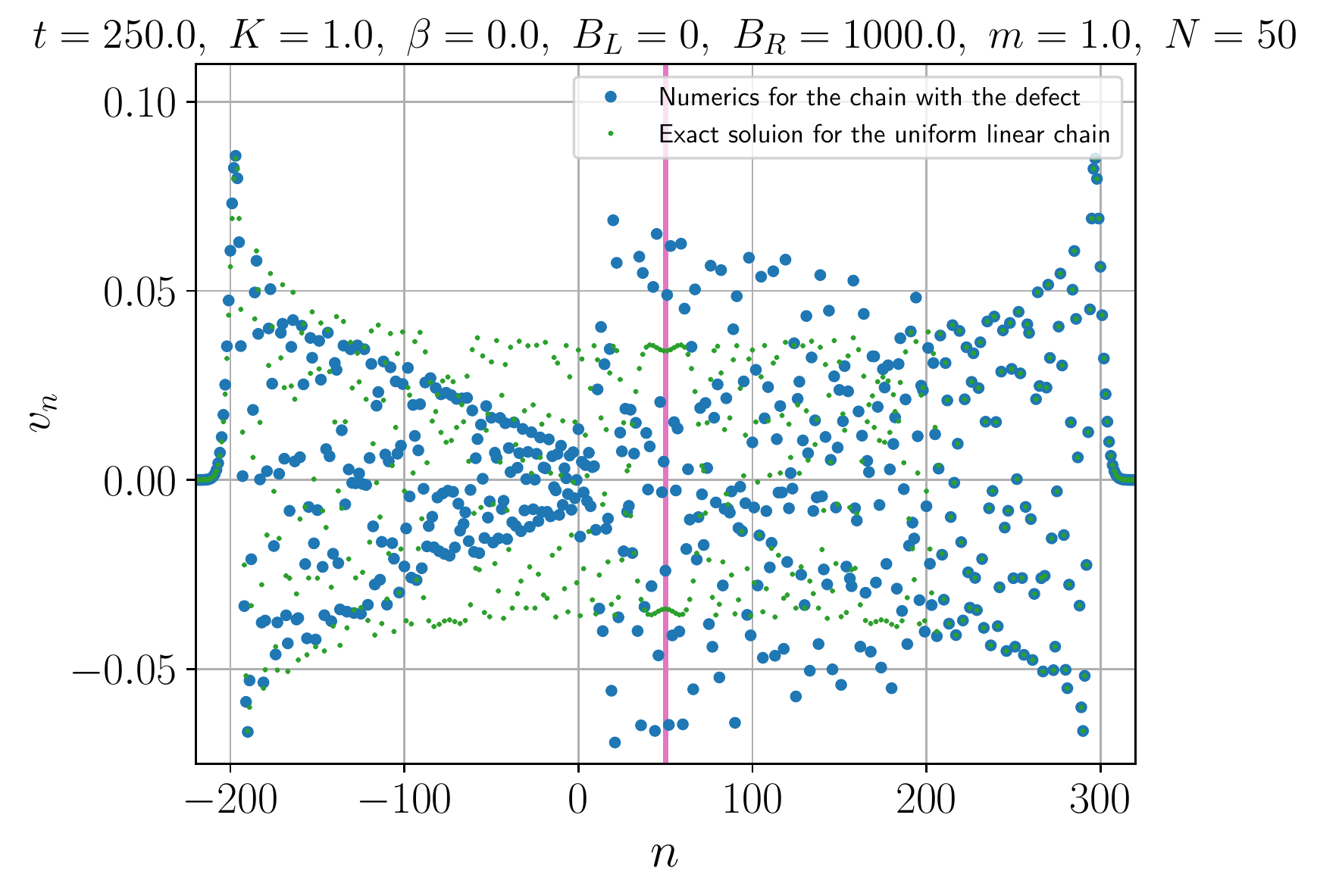}
\caption{The particle velocity $v_n$ versus the spatial variable $n$ in an infinite 
  linear chain with a hard, strongly nonlinear spring.
The source position $n=N$
is indicated
by the vertical magenta solid line. One can clearly see a shadow behind the
defect at $n=0$}
\label{V_loc_dif_non.pdf}
\end{figure}

A possibility to eliminate the localized oscillation is to consider a pulse
loading applied at some distance $N>0$ outside the defect. For such a case, in study
\cite{Gavrilov2024}, we have shown that in the linear chain 
the anti-localization is observed in some one-sided neighborhood just behind the defect, 
which expands with time. The amplitude of the localized oscillation in the
linear system exponentially decreases with $N$; therefore this oscillation
is not observable for large enough $N$.
The order of decay for the particle velocities in such a shadow zone 
is the same as it is observed when the loading is applied at a defect, i.e.,
$t^{-3/2}$. 
In Fig.~\ref{V_loc_dif_non.pdf}, the numerically found
particle velocity $v_n(t)$ versus the spatial variable $n\in\mathbb Z$ 
in an infinite linear chain, with an inclusion in the form of one
strongly nonlinear spring,
is shown for a fixed value of time. Here, we take an extremely large value
$B_R=1000$ to obtain a clearly visible effect.
For reference, the wave-field in the
uniform linear chain given by Eq.~\eqref{Sro-bessel}, 
is also displayed. 
One can compare Fig.~\ref{V_loc_dif_non.pdf} with Fig.~1 in
\cite{Shishkina2023cmat}, where 
the numerically found
particle velocity $v_n(t)$ versus the spatial variable $n\in\mathbb Z$ is
shown for the analogous problem for the linear chain with an isotopic defect. The
plots are very similar, and the shadow zones behind the defect are clearly
recognizable in both cases.

\begin{figure}[htbp]
\centering\includegraphics[width=1\columnwidth]{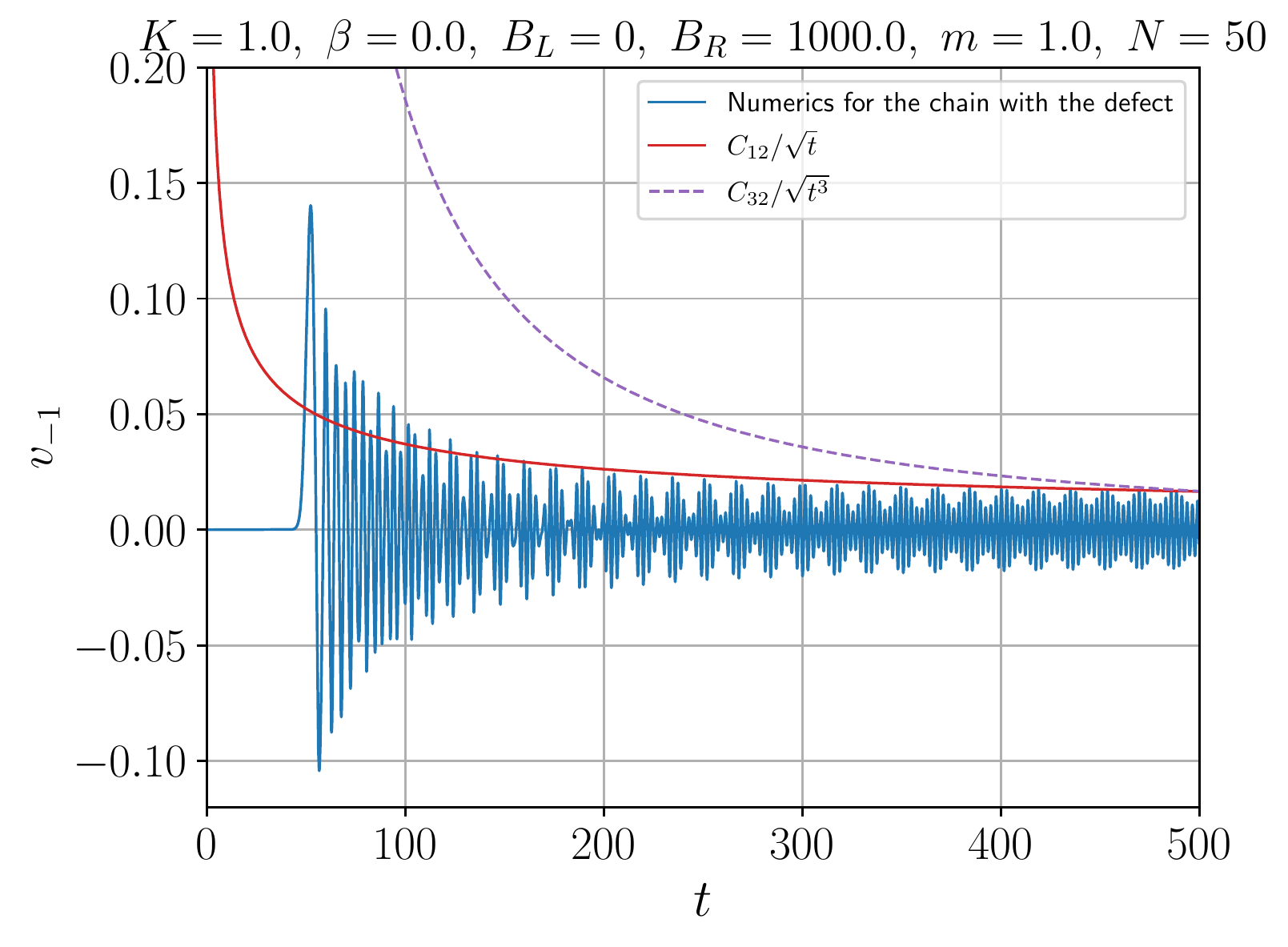}
\caption{The particle velocity $v_{-1}$ versus time in an infinite linear chain with a nonlinear spring. The order of decay can be estimated as $1/\sqrt t$}
\label{Vt_aloc_dif_non.pdf}
\end{figure}

In Fig.~\ref{Vt_aloc_dif_non.pdf} the numerically found solution for 
the particle velocity $v_{-1}(t)$ versus the time $t$ is shown for the same
system. The choice
$n=-1$ corresponds to the first particle behind the defect, which is not
connected to the defective spring.
One can see that 
the estimate for the order of the decay should be taken as $t^{-1/2}$,
not $t^{-3/2}$ as usual when the (complete) anti-localization is observed. 
Thus, we really observe a weakening of the oscillation behind the defect, but not an {\it
asymptotic} one. The leading order term of the large time asymptotics
for the oscillation amplitude is
relatively small, but not zero. 

%

\section{Conclusion}
\label{sect-conclusion}

In the paper, we have presented the numerical results for the investigation of
several strongly nonlinear problems concerning the quasi-waves propagation in an infinitely long
$\beta$-FPUT chain with a defect. Though we have demonstrated the
results, which correspond to a specific choice of the problem parameters, in
our work, we have considered a wide band of them. This allows us to claim 
that the phenomenon of anti-localization of non-stationary 
waves is not somehow related to the linearity of the system under consideration. Anti-localization is 
natural for both linear and nonlinear systems with defects.  

Depending on the parameters of the system, we have the following alternatives:
\begin{enumerate} 
\item
An anti-localized propagating component of a wave-field may co-exist with an impurity-induced
localized wave; see, e.g., Fig.~\ref{V_iso_light.pdf}, which corresponds to a light
isotopic defect. The localized mode forms a peak of the amplitude near a
defect, but this peak is observed inside a zone, where the oscillation amplitude is almost zero.
\item
An anti-localized propagating component can be
observed near a defect in the absence of impurity-induced localized modes;
see, e.g., Fig.~\ref{V_iso_heavy.pdf}, which corresponds to a heavy isotopic
defect.
\item 
The anti-localization is not observed for systems with parameters 
at the boundary of the localization domain
\footnote{A domain in the problem parameter space where the wave localization occurs.};
see, e.g., Fig.~\ref{V_boundary.pdf}, which corresponds to a mass-spring
defect.
Thus, we have shown that the absence of the anti-localization in a uniform
system is not related to the uniformity itself.  
A uniform system is just the particular case of a system with such boundary
parameters. 
\end{enumerate} 
These results were shown previously in
\cite{Shishkina2023jsv} for a continuum linear system in an analytic way.
In this paper, we have demonstrated numerically an analogous behavior of strongly nonlinear systems.

In Sect.~\ref{sect-2springs}, we have demonstrated that embedding of the point strong
nonlinearities into a linear chain can also lead to a weakening of the
propagating component of the wave-field in a neighborhood of such a defect.
Though this effect is quite similar to one which we call the anti-localization, in
the latter case, we observe weakening but not zeroing of the leading-order term. This can be called
``the incomplete anti-localization''.

Although the conclusions in the current paper are very similar to ones previously 
obtained for 
the linear case, we claim that they cannot be firmly deduced from any linear
consideration. The observations that support this point of view are as follows:
\begin{itemize} 
\item The considered nonlinearity clearly quantitatively 
  alters the solution at least everywhere excepting the neighborhood of the defect;
  see, e.g., Figs.~\ref{V_iso_heavy.pdf}, \ref{V_semi.pdf}a;
\item The considered nonlinearity clearly quantitatively alters solution at the defect 
for the finite (not very small and not very large) values of time;
  see, e.g., Figs.~\ref{V_iso_heavy_t.pdf}a, \ref{V_semi.pdf}b;
\item In the case of a heavy isotopic defect, nonlinearity qualitatively
  alters the solution at the defect for finite and large times. Namely,
  oscillation asymmetry is observed; see Fig.~\ref{V_iso_heavy_t.pdf}a.
  This effect is impossible \cite{Shishkina2023cmat} in the linear system.
\item Anti-localization can be observed in a neighborhood of a nonlinear
  impurity-induced localized mode; see, e.g., Fig.~\ref{V_iso_light.pdf};
\item The phenomenon of the incomplete anti-localization discussed in Sect.~\ref{sect-2springs} is observed for 
a nonlinear system, which is uniform in the linear approximation;
\item The anti-localization in a linear semi-infinite chain  has been
demonstrated in the strongly nonlinear case;
see Fig.~\ref{V_semi.pdf}; although
before (\cite{Jackson1978}, p.~137) this effect was explained as a specific peculiarity of a linear system;

\end{itemize} 
In our opinion, a nonlinearity, of course, could destroy the anti-localization as suggested in 
\cite{Jackson1978}.  
Indeed, 
in domains where localized waves are not observed, the
linear equations are applicable for large times. However, the initial and boundary conditions
for these equations are completely different from ones that we consider
investigating the anti-localization in linear systems. Nonlinear localized waves of a finite amplitude 
(impurity-induced ones and solitary waves at the leading wave-fronts), in
principle, could destroy the anti-localization. These
localized structures are the sources, from which the
perturbations propagate into a neighborhood of the defect. In \cite{Gavrilov2024} we have shown that 
in the linear case a source outside 
the defect destroys anti-localization inside a zone between
the source and the defect. 
The essentially nonlinear stage of the motion,
which corresponds to finite values of time, also can affect
defect's oscillation at large times, see  Fig.~\ref{V_iso_heavy_t.pdf}a.
Nevertheless, the
investigation in this paper shows that the anti-localization survives in
nonlinear systems. Thus,
the anti-localization can also be observed in weakly nonlinear systems for very large
times, i.e., in the thermodynamic limit. 

Note that introducing nonlinearity into a linear system possessing a
localized mode with the corresponding frequency inside a stop-band, in some cases, can lead to destroying the mode.
This happens at least in the case when the stop-band correspond to smaller values of frequency 
compared with the pass-band
\cite{kaplunov2008example,indeitsev2004localization}
due to the emergence of super-harmonics, whose frequencies are multiples of the localized mode frequency. These 
frequencies can leave the stop-band yielding the leakage of
energy of the localized mode.
In this paper, we consider systems with a stop-band corresponding to higher values
of the frequency compared with a pass-band. This situation is quite usual for discrete systems 
with zero on-site potential. The anti-localization is the phenomenon, which is closely related to the localization; thus, 
in nonlinear systems where a stop-band
corresponds to small values of the frequency (this is typical for the
continuous case) the anti-localization requires an additional investigation. 

\section*{Acknowledgements}

The authors are grateful to E.F.~Grekova, S.D.~Liazhkov, Yu.A.~Mochalova, A.V.~Porubov
for useful and stimulating discussions.
%
%
The research is supported by the Ministry of Science and Higher Education of the Russian Federation (project
124040800009-8).

%
%
%
%
%
%

\appendix
\section{Non-dimensionalization}
\label{AppA}
In the dimensional form, the equations of motion are:
\begin{multline}
 \minfty \frac{\d^2 {\tilde u}_n}{\d \tilde t^2}
-\Kinfty\big((\tilde u_{n+1}-\tilde u_n)-(\tilde u_n-\tilde u_{n-1})\big)
\\
-\Binfty\big((\tilde u_{n+1}-\tilde u_n)^3-(\tilde u_n-\tilde u_{n-1})^3\big)
\\
=
\Big((\tilde K-\Kinfty)\big((\tilde u_1-\tilde u_0)-(\tilde u_0-\tilde u_{-1})\big)
\\
+\big((\tilde B_R-\Binfty)(\tilde u_1-\tilde u_0)^3-(\tilde B_L-\Binfty)(\tilde u_0-\tilde u_{-1})^3\big)
\Big) \delta_n
\\
- (\tilde K-\Kinfty)\big((\tilde u_1-\tilde u_0)\delta_{n-1}-(\tilde u_0-\tilde u_{-1})\delta_{n+1}\big)
\\
- \big((\tilde B_R-\Binfty)(\tilde u_1-\tilde u_0)^3\delta_{n-1}\\
{-(\tilde B_L-\Binfty)(\tilde u_0-\tilde u_{-1})^3\delta_{n+1}\big)}
\\
-(\tilde m-\minfty)
  \frac{\d^2{\tilde u}_0}{\d \tilde t^2}\delta_n.
\label{m-s-incl-nd}
\end{multline}
Here, $n\in\mathbb Z$; $\tilde t$ is the time, $\tilde u_n(\tilde t)$ is the displacement of the particle with a number $n$;
$m_\infty$ is the mass for particles outside the defect; 
$K_\infty$, $B_\infty$ are the linear and cubic stiffnesses for the bonds outside
the defect; $\tilde m>0$ is the mass of the particle at the defect,
$\tilde K$ is the linear stiffness for the bonds at the
defect;
$\tilde B_L$ and $\tilde B_R$ are the cubic stiffnesses for the bonds to the
left and to the right of the defect;
$\de_n$ is {the Kronecker delta} ($1$ if and only if $n=0$, $0$ otherwise, {$n\in\mathbb Z$}).

The initial conditions are
\begin{equation} 
  \tilde u_n(0)=0, \quad  \frac{\d\tilde u_n}{\d \tilde t}(0)=
  \frac{\tilde p\delta_{n-N}} {m_\infty+\delta_n(\tilde m-m_\infty)}.
\label{ic-pre}
\end{equation} 
Initial conditions in the form of Eq.~\eqref{ic-pre} correspond to the pulse
loading $\tilde p\delta(t)$ applied to the particle with label $N$. 
Here 
$\tilde p$ is the magnitude of the loading,  
$\delta(t)$ is the Dirac delta-function.

Put 
\begin{equation}
\begin{gathered} 
  \omega=\sqrt{\frac{K_\infty}{m_\infty}},
  \quad
  t=\omega \tilde t,
  \quad 
  u_n=\frac{\tilde u_n \Kinfty}{\tilde p \omega},
  \\
  K=\frac {\tilde K}{K_\infty},
  \quad
  m=\frac{\tilde m}{\minfty},
  \\
  B_L=
  \frac{\tilde B_L \tilde p^2 }{\minfty \Kinfty^2},
  \quad
  B_R=
  \frac{\tilde B_R \tilde p^2 }{\minfty \Kinfty^2},
  \quad
  \beta=
  \frac{\Binfty\tilde p^2 }{\minfty \Kinfty^2}.
\end{gathered} 
\label{d-var}
\end{equation}
Using dimensionless quantities \eqref{d-var}, Eq.~\eqref{m-s-incl-nd} may be
rewritten in the following dimensionless form:
\begin{multline}
\ddot{u}_n
-(u_{n+1}-2u_n+u_{n-1})
\\-\beta\big((u_{n+1}-u_n)^3-(u_n-u_{n-1})^3\big)
\\=
\big(
  (K-1)\big((u_1-u_0)-(u_0-u_{-1})\big)
\\+\big((B_R-\beta)(u_1-u_0)^3-(B_L-\beta)(u_0-u_{-1})^3\big)
\big) \delta_n
\\
- (K-1)\big((u_1-u_0)\delta_{n-1}-(u_0-u_{-1})\delta_{n+1}\big)
\\- \big((B_R-\beta)(u_1-u_0)^3\delta_{n-1}-(B_L-\beta)(u_0-u_{-1})^3\delta_{n+1}\big)
\\
-(m-1)\ddot{u}_0\delta_n
,
\label{m-s-incl}
\end{multline}
which is equivalent to Eqs.~\eqref{baseeq1}--\eqref{baseeq2} if
notation \eqref{eq:constants} is accepted. The initial conditions 
\eqref{ic-pre} can be reformulated in the form 
of Eq.~\eqref{ic}.

\section{An infinite linear chain with an isotopic defect}
\label{App-B}
For the reader's convenience, we provide here some important formulas obtained in 
\cite{Shishkina2023cmat} that 
describes the solution
of the problem considered in Sect.~\ref{sect-isotope}, Eqs.~\eqref{eq:B-def}--\eqref{eq:N=0}
in the linear case $\beta=0$.
The exact solution for the particle
velocities in the integral form is an inverse Fourier transform:
\begin{multline}
\VA_n=
-\frac{\I}{2\pi}\left(\int_{\mathbb P}+\int_{\mathbb S}\right)
\Omega\mathscr G_n(\Omega)\pFO\EXP{-\I\Omega t} \, \d\Omega+\cc
\\
=I_n^{\mathrm{pass}}+I_n^{\mathrm{stop}}+\cc
=v_n^{\mathrm{pass}}+v_n^{\mathrm{stop}},
\label{dot-u-fourier-gen}
\end{multline}
where $\mathscr G_n(\Omega)$ is the corresponding Green function in the
frequency domain:
\begin{align}
&
\mathscr G_n(\Omega)=-\frac{
\EXP{\I  |n|\sign \Omega\, \arccos\frac{2-\Omega^2}2}}
{(\DM)\Omega^2+\I\Omega \sqrt{4-\Omega^2}}
,\ \  
\Omega\in\mathbb P;
\label{Green-function0-lower-e}
\\
&
\mathscr G_n(\Omega)=\frac{(-1)^{|n|}2^{|n|}}
{\Phi^{|n|-1}(\Omega)\big((-m\Omega^2+2)\Phi(\Omega)+4\big)}
,\ \ 
\Omega\in\mathbb S;
\label{Green-function0-upper-e}
\end{align}
$\cc$ are the complex conjugate terms.
Here,
\begin{equation}
\Phi(\Omega)\=\Omega^2-2+|\Omega|\sqrt{\Omega^2-4};
\end{equation}
\begin{equation}
\mathbb P\=[0,\Omega_\ast], \qquad
\mathbb S\=(\Omega_\ast,\infty)
\end{equation}
are the pass-band and the stop-band of the whole non-negative frequency band $\Omega\in[0,+\infty]$; 
$\Omega_\ast=2$ is the cut-off frequency which separates the bands.


\begin{remark} 
In \cite{Shishkina2023cmat} mostly the heat, i.e., the kinetic energy, transfer processes are under
consideration. Thus, the solution is obtained for the particle velocities, not
for the displacements. This is also useful because the displacements involve
a non-vanishing non-oscillating component (a rigid body mode of motion).
\end{remark} 

The integral $I_n^{\mathrm{stop}}$ 
describes, in particular, an impurity-induced non-vanishing localized
oscillation, which exists only in the case of a light defect
\begin{equation}
0<m<1. 
\end{equation}
If the last inequality is true, in the interval $(0,\Omega_\ast)$ there exists a
simple root
\begin{equation}
\Omega_0=\frac2{\sqrt{m(2-m)}}
\label{OSC-fr-root}
\end{equation}
of the denominator for $\mathscr G^{\mathrm{stop}}$.
Therefore,
integral $I^{\mathrm{stop}}$ does not exist in the classical
sense.
In the
latter case,
integral $I^{\mathrm{stop}}$ 
should be considered as the Fourier transform for
a generalized function. To regularize this, {we can apply} 
\cite{Shishkina2023jsv,Shishkina2023cmat}
the limit absorption
principle. {Finally, for $t\to\infty$ at a fixed position $n$ one gets 
\begin{gather}
v_n^{\mathrm{stop}}=I^{\mathrm{stop}}+\cc=H(1-m)L_n(t)+O(t^{-1}),
\label{c-trapped}
\\
{L}_n=\frac{2(\DM)(-1)^{|n|+1}m^{|n|-1}H(1-m)}{(2-m)^{|n|+1}}
\cos\Omega_0 t
,
\label{v-loc}
\end{gather} 
where $t\to\infty$,
see \cite{Shishkina2023cmat}.
Here $L_n(t)$ is the localized non-vanishing oscillation,
$H(\cdot)$ is the Heaviside step-function.

The phenomenon of the anti-localization is mostly related to the integral
$I^{\mathrm{pass}}$, which describes the propagating part of the wave-field.
Following to \cite{Gavrilov2022ijhmt},
we have estimated it in 
\cite{Shishkina2023cmat}
on a moving at an arbitrary sub-critical speed $w$ point of
observation. 
Taking
into account that solution 
\eqref{dot-u-fourier-gen}
clearly is an even function of $n$, put
\begin{gather}
|n|=w t,
\label{xi-w}
\\
0<w<1
\label{w-ineq}
\end{gather}
in the expression for $I^{\mathrm{pass}}$ defined by Eq.~\eqref{dot-u-fourier-gen}. The obtained integral can be
estimated for $t\to\infty$ 
using the method of stationary phase. This provides an asymptotic continuum
solution:
\begin{multline} 
v_n^{\mathrm{pass}}=
  I^\mathrm{pass}+\cc\\=\frac{{A}(w)}{\sqrt t}\cos  \Big(\phi(w)\,t+\frac{\pi}{4}+\psi(w) \Big)
+
O(t^{-3/2}),
\label{sum-I_1+I_2-single-wave}
\end{multline} 
\begin{gather}
 \begin{gathered} 
{A}(w)\=\!
\frac{H(1-w)\,w}
{\sqrt\pi{(1-w^2)^{1/4}}\big((\DM)^2(1-w^2)+w^2\big)^{1/2}},
\\
\\
\end{gathered} 
\label{sum-I_1+I_2-single-wave-amp}
\\
\phi(w)\=
2(w\arccos w-\sqrt{1-w^2} ),
\label{phi_ast-expr}
\\
\psi(w) \=
\arctan 
\frac
{(\DM)\sqrt{1-w^2}}
{w}.
\label{psi}
\end{gather}
Note that for $w>1$
there is no stationary point, and the corresponding term of order $1/\sqrt t$ is
zero. This fact is taken into account by introducing the multiplier $H(1-w)$ in
the numerator of the right-hand side of 
Eq.~\eqref{sum-I_1+I_2-single-wave-amp}. 
In any system with an isotopic impurity,
i.e., in the case \eqref{m-def}, we have:
\begin{equation}
  {A}(w)=\frac{\sqrt 2 w}{\sqrt{\pi}|m-1|}+O(w^3),\quad w\to0;\qquad A(0)=0;
\label{A-as}
\end{equation}
and, thus, the amplitude $A(w)$ 
of the propagating part $v^{\mathrm{pass}}$
for the particle velocities
is small  in a certain expanding (since $w=|x|/t$)
neighborhood of the impurity. 
{\it We call this phenomenon the anti-localization of non-stationary
waves.}
Finally, 
the asymptotic expansion for the right-hand side of 
Eq.~\eqref{dot-u-fourier-gen}
{at $n=0$ (just at the
impurity)} has the following form
\begin{multline}
v_0(t)=H(M-K)L_0(t)+
\frac{\cos\left(2t-\frac{3\pi}4\right)}{2\sqrt\pi(\DM)^2\, t^{3/2}}
\\+ o(t^{-3/2})
.
\label{rubin-f}
\end{multline}
The first term  in the right-hand side of 
Eq.~\eqref{rubin-f} is the contribution from the poles
$\pm\Omega_0$ and describes localized oscillation, which exists if and only
if $m<1$. 
The second term is the anti-localized part of the wave-field at the inclusion
expressed as 
the total contribution from the cut-off frequency
$\Omega_\ast$ for both integrals $I^{\mathrm{stop}}$ and $I^{\mathrm{pass}}$.

In the absence of the impurity, the anti-localization is not observed:
\begin{multline}
  A(w)=\frac1{\sqrt{\pi{\sqrt{1-w^2}}}}=\frac1{\sqrt\pi}+O(w^2), \quad w\to0;
  \\\qquad A(0)\neq0.
  \label{A(0)}
\end{multline}
Though the last formula can be formally derived from 
Eq.~\eqref{sum-I_1+I_2-single-wave-amp}, this way is incorrect since Eq.~\eqref{sum-I_1+I_2-single-wave-amp}
is obtained assuming Eq.~\eqref{m-def}. The correct way to obtain formula \eqref{A(0)} is to use 
the exact solution 
\eqref{Sro-bessel}  \cite{Gavrilov2022ijhmt}.
Thus, the anti-localization in the linear system is an asymptotic weakening of
the propagating component of the wave-field. The order of decay for the vanishing component
is $1/\sqrt{t}$ in the uniform system and $1/\sqrt{t^3}$ in the system with a
defect (near the defect).

\section{A semi-infinite linear chain}
\label{AppC}
The exact solution is \cite{Liazhkov2025}
\begin{equation}
v_n=J_{2(n-1)}(2t)+J_{2n}(2t)
, \qquad
n\in\mathbb N.
\label{semi-exact}
\end{equation}

The anti-localization in the semi-infinite linear chain ($\beta=0$)
was demonstrated in study \cite{Gavrilov2023DD}, where the asymptotic solution
for $v_n$ in the form of the right-hand side of Eq.~\eqref{sum-I_1+I_2-single-wave} wherein
\begin{gather}
{A}(w)=\frac{2 w}{\sqrt{\pi }\sqrt[4]{1-w^2}},
\\
\tphi
=
\arctan 
\frac
{\sqrt{1-w^2}}
{w}
\label{psi-def},
\\
w=\frac{|n-1|}t
\end{gather}
was obtained. The expression for $\phi$ coincides with 
\eqref{phi_ast-expr}. Clearly, $A(0)=0$, which implies the anti-localization
effect. 
For the end particle, one has the anti-localized solution \cite{Gavrilov2023DD}:
\begin{gather} 
  v_1=\frac{\sin(2t-\frac\pi4)}{\sqrt\pi t^{3/2}}+o(t^{-3/2}).
\end{gather} 

\bibliographystyle{apsrev4-2}
\bibliography{bib/serge-gost,bib/all}

\end{document}